\documentclass{article}

\usepackage{arxiv}

\usepackage{times}
\usepackage{epsfig}
\usepackage{graphicx}
\usepackage{amsmath}
\usepackage{amssymb}
\usepackage{caption}
\usepackage{subcaption}
\usepackage{booktabs} 
\usepackage{wrapfig}
\usepackage{url}

\usepackage[dvipsnames]{xcolor}
\usepackage[most]{tcolorbox}

\title{\LARGE \bf MAAD: A Model and Dataset for \\ ``Attended Awareness'' in Driving}

\usepackage{bbm}

\def\LLL{\mathcal{L} }

\author{
Deepak Gopinath\thanks{Department of Mechanical Engineering, Northwestern University. $^{\dagger}$Toyota Research Institute, Cambridge, MA. $^{\ddagger}$Woven Planet, Tokyo, Japan. Presented at the Ninth International Workshop on Egocentric Perception, Interaction and Computing: EPIC@ICCV2021, Virtual Meeting.}, Guy Rosman$^{\dagger}$, Simon Stent$^{\dagger}$, Katsuya Terahata$^{\ddagger}$ \And
Luke Fletcher$^{\dagger}$, Brenna Argall$^{*}$, John Leonard$^{\dagger}$
}
\date{}
\begin{document}
\maketitle

\begin{abstract}
We propose a computational model to estimate a person's attended awareness of their environment. We define ``attended awareness'' to be those parts of a potentially dynamic scene which a person has attended to in recent history and which they are still likely to be physically aware of. 
Our model takes as input scene information in the form of a video and noisy gaze estimates, and outputs visual saliency, a refined gaze estimate and an estimate of the person's attended awareness. 
In order to test our model, we capture a new dataset with a high-precision gaze tracker including 24.5 hours of gaze sequences from 23 subjects attending to videos of driving scenes. The dataset also contains third-party annotations of the subjects' attended awareness based on observations of their scan path.
Our results show that our model is able to reasonably estimate attended awareness in a controlled setting, and in the future could potentially be extended to real egocentric driving data to help enable more effective ahead-of-time warnings in safety systems and thereby augment driver performance. 
We also demonstrate our model's effectiveness on the tasks of saliency, gaze calibration and denoising, using both our dataset and an existing saliency dataset.
We make our model and dataset available at 
\url{https://github.com/ToyotaResearchInstitute/att-aware/}.
\end{abstract}

\section{Introduction} \label{sec:intro}
\setcounter{footnote}{0} 
We define ``attended awareness'' to be those parts of a potentially dynamic scene which a person has attended to in recent history and which they are still likely to be physically aware of.
While attended awareness is difficult to objectively measure, in certain situations such as driving, it is possible to infer, at least to some useful degree. Driving instructors routinely assess a driver's behavior based on their estimated attended awareness of a given driving scene, and provide real-time feedback to ensure safety.
\begin{figure}
    \centering
    \includegraphics[width=\linewidth]{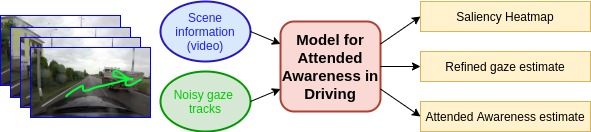}
    \caption{\textbf{Model for Attended Awareness in Driving (MAAD) overview.} Our model takes as input a video of a scene, as seen by a person performing a task in the scene, along with noisily registered ego-centric gaze sequences from that person. The model estimates (i) a saliency heat map, (ii) a refined gaze estimate, and (iii) an estimate of the subject's attended awareness of the scene. We evaluate our model using a unique annotated dataset of third-person estimates of a driver's attended awareness. By explicitly estimating a person's attended awareness from noisy measurements of their gaze, MAAD can improve human-machine interactions. In the driving example, such interactions might include safety warnings in situations where attended awareness is deemed insufficient.}
    \label{fig:overview}
\end{figure}
In the context of human-machine interaction, inferring human attended awareness in a given scenario is valuable for machines to facilitate seamless interaction and effective interventions. Human gaze information can be used by machines for this purpose and many models have been developed to relate scene understanding and overt attention as saliency estimates \cite{borji2012state,bruce2016deeper,cornia2018predicting,huang2015salicon,pan2016shallow,pan2017salgan} and objectness measures \cite{alexe2012measuring,borji2015salient}. 
However, further exploration of the link between visual attention and processes of scene understanding and decision making has been constrained by a limited ability to reason about such cognitive processes, which is difficult in general contexts
\cite{salvucci2001integrated,tatler2010yarbus,yarbus1967eye}.

In this work, we propose a model to estimate attended awareness as a spatial heatmap.
The input to our model is a video of the scene under observation, and a noisy estimate of the person's gaze. Our model allows us to \textit{(i)} compute visual saliency for the scene, \textit{(ii)} leverage this to refine the noisy gaze estimate, and \textit{(iii)} combine this information over time to infer the person's attended awareness. The construction of our model is driven by axiomatic considerations that define an image translation network from the scene image to the saliency and awareness heat maps. Within a larger human-machine interaction scenario, the model becomes useful as an inference engine for human awareness enabling seamless interaction (for e.g., a person operating a semi-autonomous car or a smart wheelchair). 
We adopt a data-driven approach (as opposed to more limited-scope analytical models \cite{kircher2013driver}) to allow for scalable and more comprehensive modeling of attended awareness that can potentially rely on supervision from multiple sources (such as resulting actions/self-reporting).

\textbf{Contributions} 1) We propose a learned model that affords estimation of attended awareness based on noisy gaze estimates and scene video over time. 2) We further demonstrate how the model affords saliency estimation as well as the refinement of a noisy gaze signal. 3) We present a new dataset that explores the gaze and perceived attended awareness of subjects as they observe a variety of driving and cognitive task conditions. While the dataset is captured via a proxy hazard awareness task rather than through real or simulated driving, it serves as a useful starting point to study visual saliency and awareness in driving scenes.
\bibliographystyle{unsrt}

\section{Related Work}
\label{sec:related}

Our work builds on prior work in visual saliency estimation, situational awareness, and driving-specific exploration. We briefly summarize related work from these areas.

\textbf{Visual saliency.}
Vision scientists have long sought to model and understand the mechanisms behind our allocation of attention in a visual scene \cite{yarbus1967eye}.
Visual salience is a function of many factors, including the spatio-temporal nature of the stimulus itself as well as its relationship to neighboring stimuli and the nature of the visual system perceiving it \cite{itti1998model}. The evolution of computational approaches to estimating visual image salience has mirrored that of other popular computer vision problems, with largely hand-engineered models---designed to mirror certain bottom-up and top-down attention processes in the human visual system (e.g.~\cite{borji2012probabilistic,itti1998model})---giving way to supervised, data-driven approaches (e.g.~\cite{huang2015salicon,judd2009learning}). 
Unlike image saliency, video saliency models consider spatial as well as temporal information to detect objects of interest in a dynamic way.

Spatio-temporal saliency estimation opens inquiry into how processes of visual attention, situational awareness and task-related decision making are connected. Previous attempts have been made to computationally model situational awareness (see e.g.~\cite{bellet2012computational,mccarley2002,wortelen2013dynamic}). Our approach to modeling situational awareness is unique in that we try to explicitly estimate the parts of the visual scene to which a person has attended using a spatio-temporal model for gaze and scene understanding. 
The three stages of forming situational awareness consists of perception, comprehension and projection~\cite{endsley1995toward} and in
our work we focused primarily on perception. We aim to model, from noisy observations of a person's scan path, the set of objects and scene structures which that person is likely to have attended to, and therefore might be better able to incorporate into their future decision-making. 
While we note that peripheral vision alone can achieve visual awareness in many settings~\cite{wolfe2017more}, we focus on objects of fixation, since we are concerned primarily with estimating when drivers fail to notice potential driving hazards, which are known to strongly induce fixations~\cite{crundall2012some}.

\textbf{Data-Driven Saliency Modeling and Datasets.}
Data-driven approaches to image and video saliency rely on state-of-the-art deep learning architectures: CNNs \cite{bruce2016deeper,huang2015salicon,pan2016shallow}, GANs \cite{pan2017salgan} and LSTMs \cite{bazzani2017recurrent,cornia2018predicting,wang2018deep}, and single image inputs through to multi-stream inputs incorporating video, along with optical flow, depth or semantic segmentation estimates \cite{lang2012depth,palazzi2018predicting} and even additional modalities such as audio \cite{Tsiami_2020_CVPR}. In our work, similar to \cite{min2019tased}, we adopt a 3D CNN--based approach, due to its simplicity and success on other video understanding tasks such as action recognition. 

While the majority of saliency datasets explore image and video saliency under controlled viewing conditions (e.g.~\cite{mit-tuebingen-saliency-benchmark,mathe2014actions,wang2018deep}), in recent years, numerous ego-centric video gaze datasets have been developed in which subjects perform tasks as varied as cooking and meal preparation \cite{li2015delving}, playing video games \cite{borji2012probabilistic} and driving \cite{palazzi2018predicting}, in parallel to developments in applications of saliency (see, e.g.~\cite{sugano2010calibration,rahman2020classifying,cazzato2020look}. On the other hand in the driving domain, in-lab data collection procedures have been extensively adopted as they have the advantages of high accuracy, repeatability and the ability to focus on rare scenarios such as critical situations and accidents~\cite{xia2018predicting,fang2019dada, deng2019drivers,baee2019medirl}, where understanding human perception can inform safety systems approaches. 

Our dataset uses an in-lab data collection paradigm, however it differs from prior work for several reasons. 
Firstly and most notably, we capture multiple subjects observing the same visual stimuli under \textit{different cognitive task modifiers}.
Our dataset therefore allows for reasoning about the effect of different cognitive task modifiers on the visual gaze patterns, given identical visual input. 
Secondly, we provide annotations for third party estimates of a subject's attended awareness, based on observations of their scan path. For this purpose, we devise a novel reproducible annotation scheme. 
Finally, as our dataset is gathered using a high precision gaze tracker with a chin rest, the precision of the scan paths is extremely high when compared to that of eye-glass gaze tracking ego-centric datasets such as \cite{palazzi2018predicting}.

\textbf{Driving specific applications.} 
Driving is a predominantly visual task. 
The first studies into driver attention and eye scanning patterns date back over half a century \cite{senders1967attentional}. 
Since then, certain driving behaviors have been well established and modelled, such as the ``perceptual narrowing'' effect in which drivers increasingly fixate on the road ahead as task demands increase (e.g. through greater speed, increased traffic or lack of familiarity with a route) \cite{engstrom2005effects}, or the benefits of understanding driver attention when predicting a driver's future intent \cite{doshi2009,wu2019gaze}.
However, to the best of our knowledge, no models exist with the purpose of quantitatively estimating a driver's spatial awareness of a scene.
In recent years, works such as \cite{palazzi2018predicting,xia2018predicting} have used high-precision gaze trackers to create video gaze datasets both in the car and in the lab, allowing for data-driven approaches to modelling driver attention.
While we make use of the road-facing data from \cite{palazzi2018predicting} in our experiments, our model differs in one key respect. Rather than estimating visual saliency from video alone, we demonstrate how, given access to a noisy gaze estimate of a driver
it is possible to simultaneously estimate scene saliency, a denoised gaze signal and an estimate of the driver's overall awareness of the scene.

\section{Method}
\label{sec:method}
\begin{figure*}
    \centering
    \includegraphics[width=0.95\linewidth]{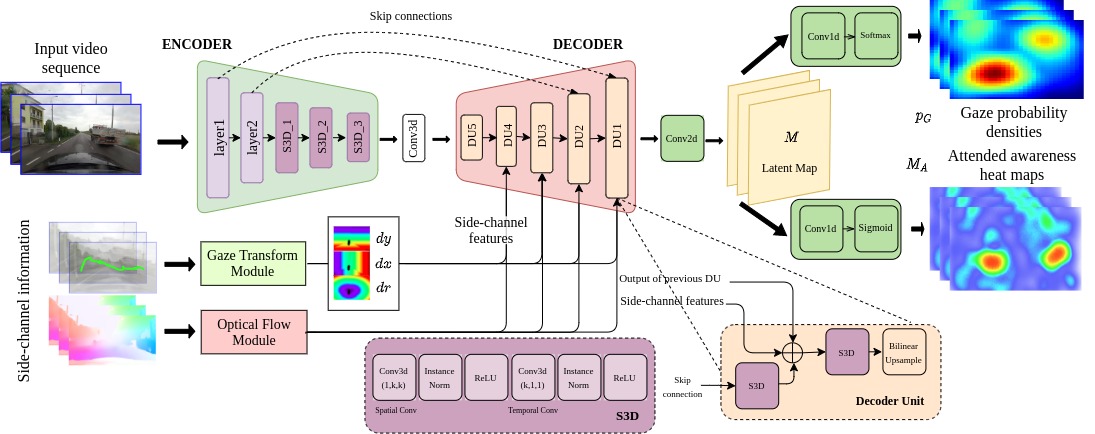}
    \caption{\textbf{Model architecture.} During training, our model takes as input a video sequence with associated scan paths from a gaze tracker, along with optical flow estimates. The video sequence is processed using a 3D convolutional encoder/decoder, and information from gaze, and optic flow is transformed and injected during the decoding stage. The model outputs two heat maps: a gaze probability density heat map, $p_G$, which provides a clean estimate of the noisy gaze input, and an attended awareness heat map, $M_A$, which provides an estimate of the driver attended awareness, both at the final frame of the input video sequence. For further details, see Section~\ref{sec:method:model}.}
    \label{fig:model}
\end{figure*}
The computational model we propose is guided by several assumptions related to human attended awareness and its relation to gaze patterns (Section~\ref{sec:method:assumptions}). These assumptions are implemented through a mixture of explicit objectives and behavioral regularizations (Section~\ref{sec:method:loss}).

\subsection{Assumptions and Supervisory Cues}
\label{sec:method:assumptions}
We use several assumptions about gaze patterns and attended awareness to define the priors in training our model:
\begin{itemize}
    \item \textbf{A1 Saliency}: Gaze tends to focus on specific regions \cite{borji2012state}, both object and stuff \cite{kirillov2019panoptic}.
    \item \textbf{A2 Attended awareness}: People tend to become aware of the objects they look at \cite{tatler2001characterising,chun2011visual}. Their attention is, however, limited in its capacity. 
    \item \textbf{A3 Awareness decay}: Awareness of an object can decrease (due to forgetting), or increase (when looking at an object) at different rates \cite{ricker2010loss}.
    \item \textbf{A4 Regularity of gaze, awareness}: Gaze and awareness maps should be regular and smooth unless otherwise warranted \cite{brady2013probabilistic,kaiser2015real}. 
    \item \textbf{A5 Awareness and motion}: As an observer moves through a dynamic scene, their awareness moves along with objects and regions in the scene and exhibits temporal persistence. \cite{makovski2009role,boon2014updating}.
\end{itemize}

\subsection{Model}
\label{sec:method:model}
We define the \textit{Model for Attended Awareness in Driving (MAAD)}, shown in Figure~\ref{fig:model}, as a fully convolutional image encoding-decoding network with shortcut links such as U-Net \cite{ronneberger2015u}. Sensor images are encoded and decoded into a latent feature map, $M(x,t)$, from which two convolutional modules emit the estimated gaze distribution $p_G(x,t)$, and the attended awareness image $M_A(x,t)$.

The gaze distribution is normalized as a probability density function (via a softmax operator). We note that $p_G(x,t)$ is a unified notation for gaze probability maps with and without a noisy gaze input from an external gaze tracker. 
In the absence of a noisy gaze input, $p_G(x,t)$ is a probabilistic form of saliency. In the rest of the paper we use $p_G(x,t)$ to denote both forms
to simplify the notation. The individual modules in the decoder are fed additional information: the (noisy) driver gaze measurement over time in the form of Voronoi maps, and optical flow maps~\cite{palazzi2018predicting} encoded as two feature layers for horizontal and vertical motion. 

\subsection{Loss Function Design}
\label{sec:method:loss}
At training time, the gaze and attended awareness maps are used to compute several supervisory and regularization terms, whose design is guided by the assumptions in Section~\ref{sec:method:assumptions}.

\subsubsection{Supervisory Terms}   
\label{sec:method:supervisory}
\textbf{Gaze Prediction.} We want $p_G$ to predict a subject's gaze as accurately as possible. This is encouraged via the primary data term:
\begin{align}
\LLL_{\text{G}} = -\sum_t\sum_{x \in X_{g}(t)} \text{log}~p_{G}(x, t),
\end{align}
where $X_g(t)$ are the 2D ground truth gaze points at time $t$.

\textbf{Perceived Awareness.} We include a supervisory source for attended awareness estimation. This term surrogates awareness estimation training by an attended awareness estimate. One approach to obtain an attended awareness estimate is to provide a person with the gaze estimate of the driver overlaid on the road scene and query how aware the driver is of specific locations in the image at particular points in time. This is further described in Section~\ref{sec:data}. The cost term reads:
\begin{align}
    \LLL_{\text{ATT}} = \sum_{(x,t) \in \textbf{labeled}} \left(M_A(x,t)-L_A(x,t)\right)^2,
\end{align}
where the summation is over all annotated samples in location $x$ at time $t$, and $L_A$ denotes the annotated measure of awareness in the range $[0,1]$ as described in Section~\ref{sec:data:annotations}.

\textbf{Awareness of Attended Objects.} Based on  (A2), we add a term that encourages awareness to be high when a person is gazing at a scene location:
\begin{align}
\LLL_{\text{AA}} = \sum_t \sum_{x \in X_{g}(t)} (M_{A}(x, t) - 1)^2.
\end{align}
    
\subsubsection{Regularization Terms}
\label{sec:method:regularization}
\textbf{Spatial Smoothness.} Based on (A4-5), we added regularity terms to both the awareness and gaze maps:
\begin{align}
 \LLL_{\text{S}, \cdot}=\int \frac{|\nabla \phi|^2}{\sqrt{|\nabla I|^2+\epsilon}}dx,
\end{align}
\vspace{-0.1cm}
where $\phi$ is $M_A$ and $p_G$ for $\LLL_{\text{S,A}}$ and $\LLL_{\text{S,G}}$ respectively and the integral is computed over all pixels.
This regularization is a variant of anisotropic diffusion \cite{perona1990scale,paris2009fast} with cross-diffusivity based on the scene image $I$.


\textbf{Temporal Smoothness.} Based on (A3-5), we apply also temporal regularization for awareness:

In order to make the map temporally consistent with respect to the locations and not just object boundaries, we use a smoothness / decay term based on the image optical flow:
\begin{flalign}
 \label{eq_temporal_smoothness}
 &\LLL_{\text{T}} = \sum_{x,t}\begin{matrix}f_{w_\text{OF}}\left(M_A(x+v_\text{OF}(x),t+1), M_A(x,t)\right) \end{matrix},\\
 & f_{w_\text{OF}}(a,b) = c_1((a-w_\text{OF}b)_{+})+c_2((a-w_\text{OF}b)_{-})^2,\nonumber
\end{flalign}
where $v_\text{OF}(x)$ is the optical flow computed on the input images and $w_\text{OF}<1$ is a weight factor that is close to 1. $()_{+}$ and $()_{-}$ denote positive and negative change respectively in awareness values when going from $t$ to $t+1$. 
Particularly, $f_{w_\text{OF}}(a,b)$ is set to be an \textit{asymmetric} loss function that penalizes awareness that increases instantaneously when attending to an object less compared to awareness that decreases rapidly via forgetting. This is accomplished by having the forgetting term, $()_{-}$, to be quadratic.

\textbf{Awareness Decay.} Based on (A3), we expect the level of awareness to decay over time, which we model via:
\begin{align}
    \LLL_{\text{DEC}} = \sum_{x,t} \left(\begin{matrix}\left(1-\epsilon_\text{DEC}\right)M_A(x,t) \\- M_A(x,t+1)\end{matrix}\right)^2,
\end{align}
where $\epsilon_\text{DEC}$ is a decay factor. 

\textbf{Attention Capacity.}  Based on (A2), we expect that the cognitive resources available to the driver do not change over time. This assumption captures the fact that the overall awareness should be similar between frames on average. 
\begin{align}
    \LLL_{\text{CAP}} = \sum_{t}\Big(\sum_{x}M_A(x,t)-\sum_{x}M_A(x,t+1)\Big)^2.
\end{align}

\textbf{Block-level consistency.}
We denote by $M_A(x,t;t_1)$ the awareness estimate at $(x,t)$ that is emitted based on the training snippet started at $t_1$, similarly for $p_G$. We define a consistency term~\cite{sajjadi2016regularization} between consecutive estimates via the loss:
\begin{align}
 \LLL_{\text{CON},\cdot} = \sum_{t_1,t}\sum_{x}\Big(\phi(x,t;t_1)-\phi(x,t;t_1+1)\Big)^2,
\end{align}
where $\phi$ is $M_A,p_G$ for $\LLL_{\text{CON},A},\LLL_{\text{CON},G}$ respectively. $\LLL_{\text{CON},\cdot}$ helps to minimize the difference between
the predictions of multiple passes of samples at the same timestamps through the network. The overall training loss is a linear combination of the previously described loss terms. See supplementary material for additional details. 
\subsection{Training Procedure}
We trained our model using PyTorch 1.7 using NV100 GPUs. 
The training was carried out using video input subsequences of length 4 seconds, sampled at 5Hz with frames resized to 240$\times$135. Our model was trained using Adam optimizer~\cite{kingma2015adam} with a learning rate of $5 \times 10^{-3}$ and a batch size of 4. The model weights were updated after the gradients were aggregated for a fixed number of steps in order to reduce the variance in the loss. The batch aggregation size was set at 8. The first two layers of the encoder (kept frozen during training) are taken from a pretrained Resnet18 implementation in Torchvision~\cite{10.1145/1873951.1874254}. Later spatio-temporal encoding is done by three layers of separable 3D convolution modules denoted as S3D in Figure~\ref{fig:model} \cite{xie2018rethinking}.
To provide optical flow estimates, we used RAFT pre-trained on FlyingChairs and FlyingThings datasets \cite{teed2020raft}. 

The decoder consists of stacked decoder units each of which receives input from three sources a) side-channel information b) skip connections from the encoder layer and c) the output of the previous decoder unit, when available. Each decoder unit consists of two submodules: 1) The skip connections are first processed via an S3D module whose output is then concatenated (channel-wise) with the side-channel information and the output of the previous decoder unit. This concatenated input is processed by another S3D module followed by bilinear upsampling that brings the output to the proper resolution for the next decoder unit. 

The decoder emits a latent map $M$ which is subsequently processed by two separate convolutional models to emit a gaze probability map, $p_G$ and an awareness heatmap denoted as a $M_A$. The softmax in the gaze convolutional module ensures that the gaze probability map is a valid probability distribution. More details regarding network architecture and training to be found in supplementary material.
\section{Dataset Description}
\label{sec:data}

Our complete dataset comprises approximately 24.5 hours of gaze tracking data captured via multiple exposures from different subjects to 6.2 hours of road-facing video drawn from the DR(eye)VE dataset \cite{palazzi2018predicting}. We concentrate our gaze capture on repeated exposures of downtown (as opposed to highway and rural) driving scenes ($77\%$) and daylight scenes ($90\%$), since these contain the most diverse visual scenarios. While the original DR(eye)VE dataset captured and registered gaze to the road-facing video using a driver head--mounted eye tracker and feature-based matching for homography estimation, accumulating significant measurement errors, we opted for an in-lab experiment. In-lab experiment offers several advantages such as higher accuracy and repeatability across subjects and cognitive task conditions. Furthermore, models trained on in-lab data has already been shown to be effective when tested on in-the-wild data \cite{xia2018predicting}. We measure gaze to an extremely high precision ($0.15^{\circ} - 0.50^{\circ}$ typical accuracy, $0.01^{\circ}$ RMS resolution, $0.05^{\circ}$ microsaccade resolution) using a tower-mounted SR Research EyeLink 1000 Plus tracker. A main novelty of the dataset is that in addition to high precision gaze, we supplement the dataset with third party annotations of attended awareness. The annotation procedure is outlined in Section 4.2. We recruited 23 subjects (aged 20-55), who each watched a subset of video clips with their heads mounted in a chin-rest after a 9-point calibration procedure. The subjects all carried US driving licenses and had at least two years of driving experience.
Their primary task was to monitor the driving scene as a safety driver might monitor an autonomous vehicle. While not a perfect substitute for in-car driving data collection, this primary task allowed for the capture of many of the characteristics of attentive driving behavior. In order to explore the effect of the cognitive task difference (vs. in-car data) on the gaze and awareness estimates, subjects viewed the video under different cognitive task modifiers, as detailed in Section~\ref{sec:data:conditions} (data collected with non-null cognitive task modifiers comprise $30\%$ of total captured gaze data). Around $45\%$ of video stimuli were watched more than once, of which $11\%$ (40 minutes) was observed by 16 or more subjects.

\begin{figure}[t]
    \centering
    \includegraphics[width=0.9\linewidth,trim={0 9cm 0 5cm},clip]{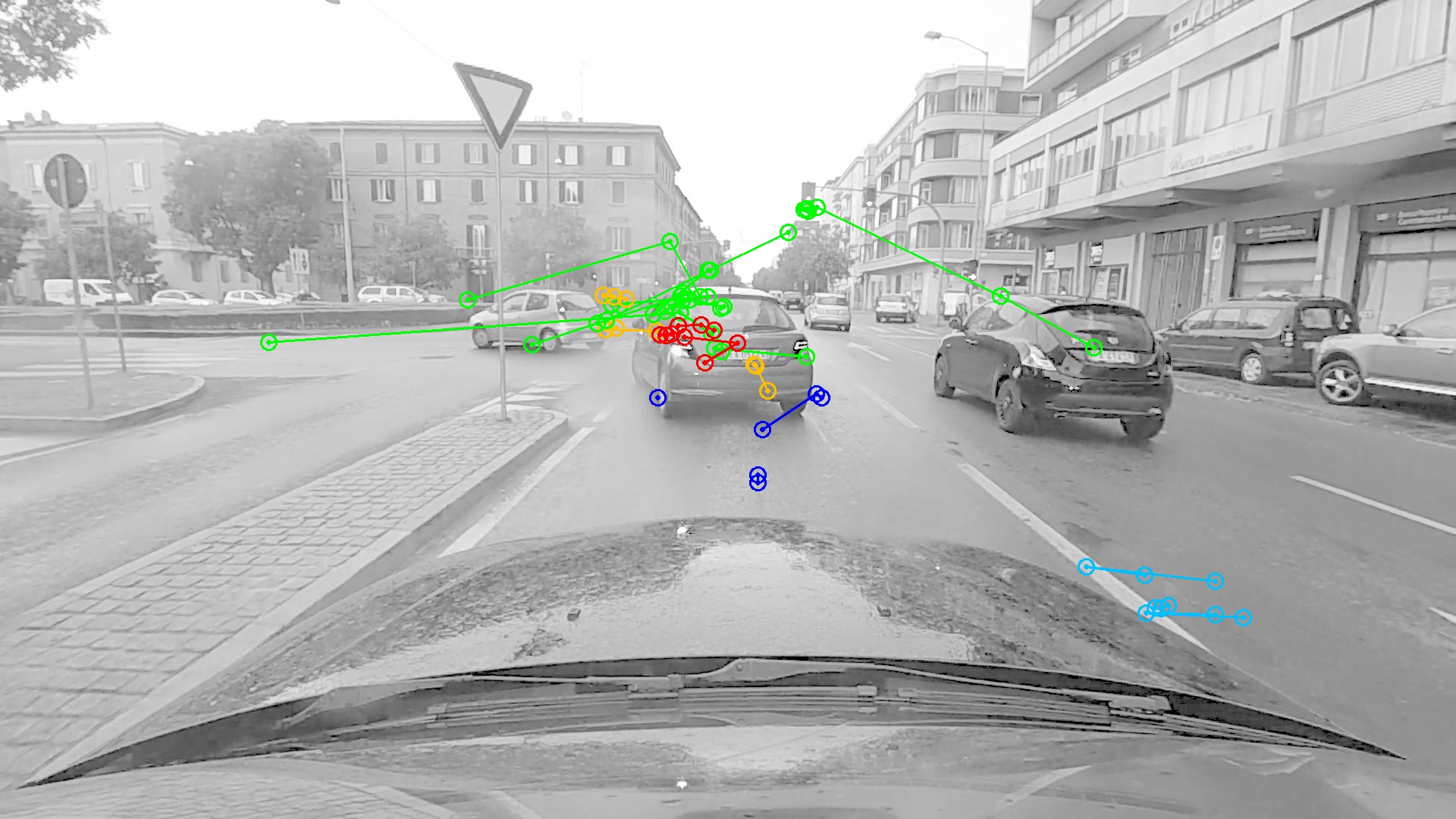}
    \includegraphics[width=0.9\linewidth,trim={0 9cm 0 5cm},clip]{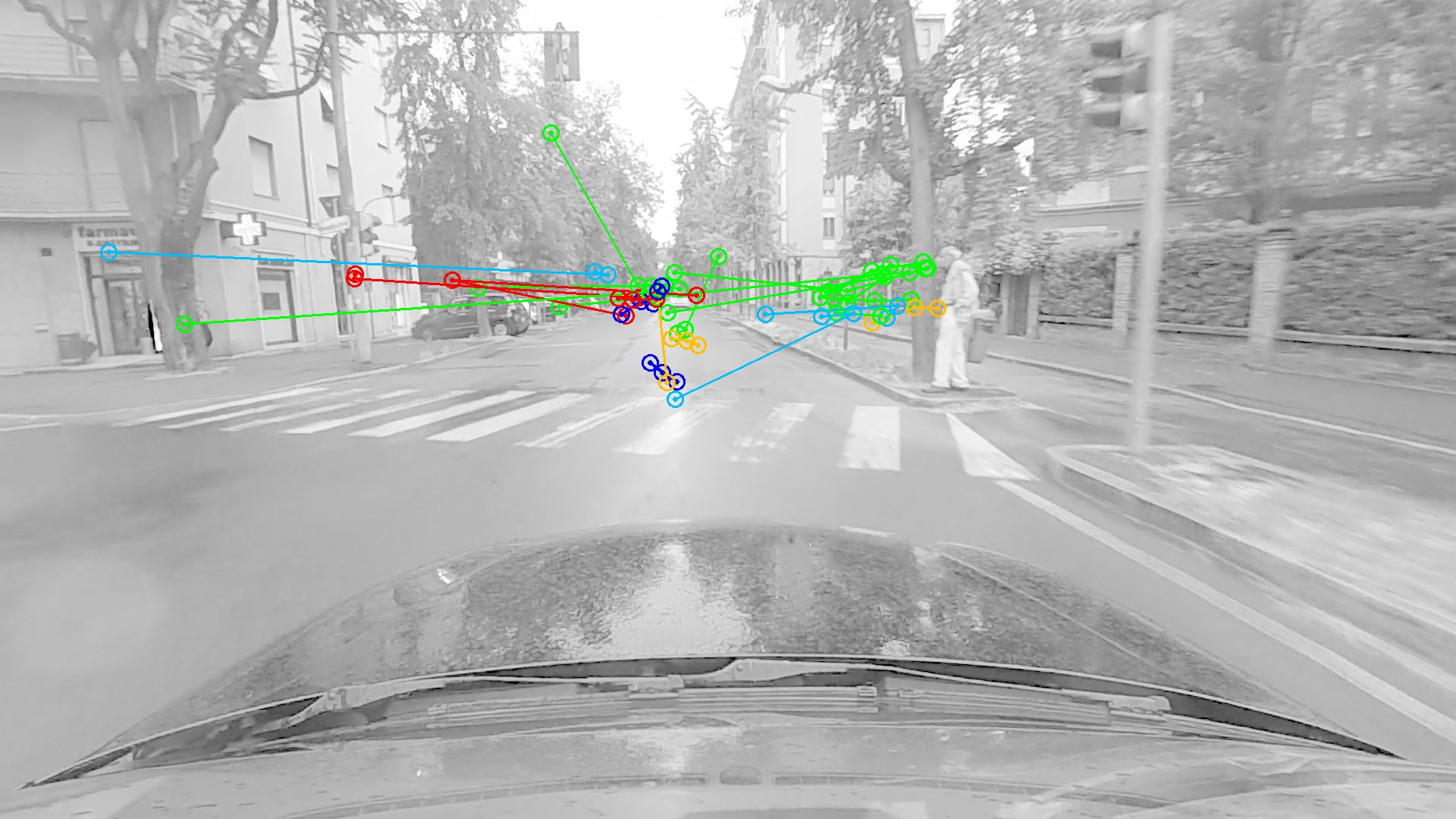}
    \includegraphics[width=0.9\linewidth,trim={0 9cm 0 5cm},clip]{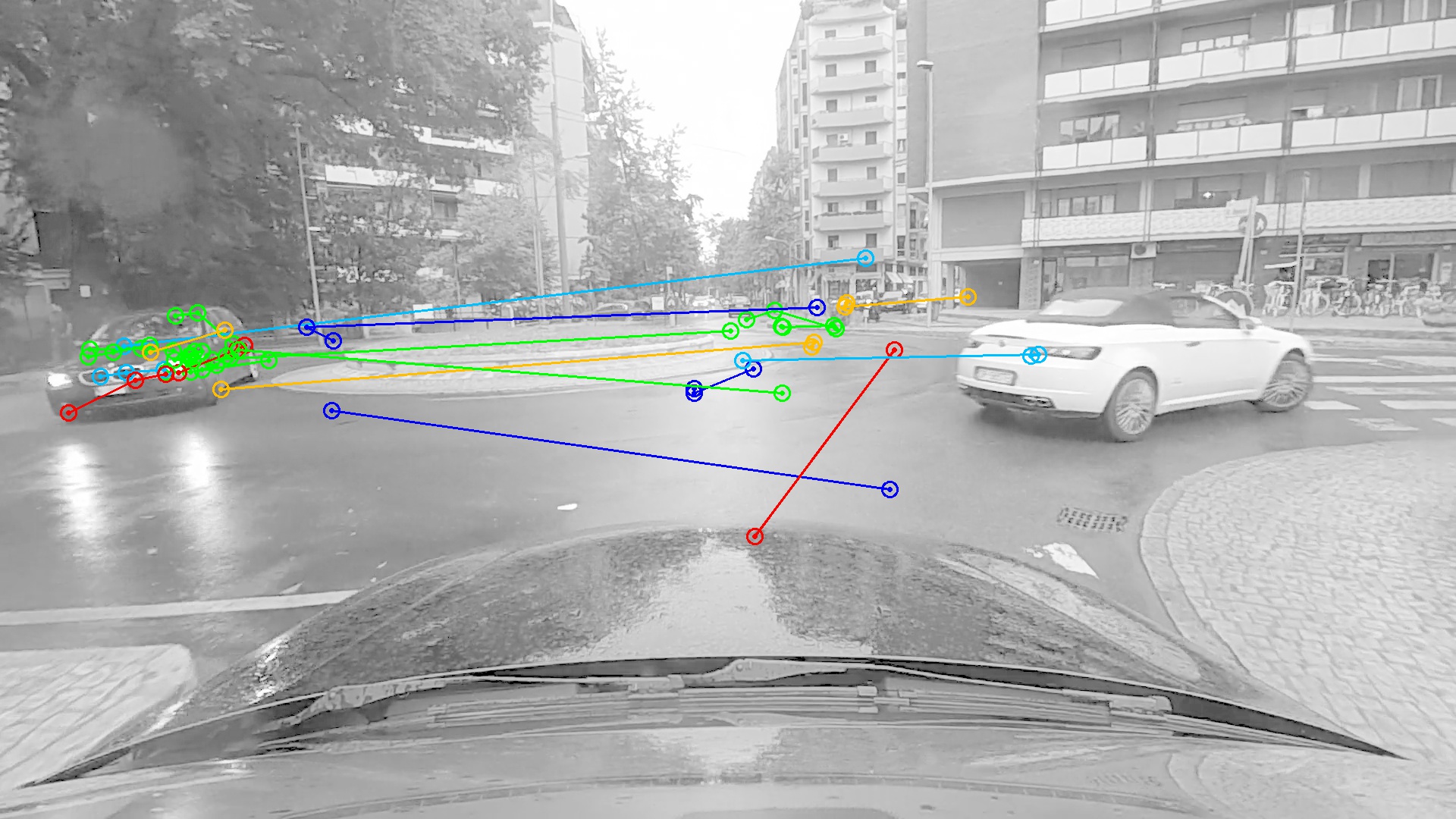}
    \caption{\textbf{What is the driver aware of?} Our dataset is unique in containing repeated views of the same driving video by 23 subjects performing the same proxy supervisory task but under different task modifiers: (i) null (green), i.e. with no secondary condition, (ii) with blurred video (yellow), (iii) with vertically flipped video (red), (iv) fixating on the road only (dark blue), and (v) reading occasional on-screen text (light blue) (see Section~\ref{sec:data:conditions} for more details). Here we overlay a sub-sampled one second gaze history per subject for a given frame in a video sequence. From observation of the scan paths of subjects, it is evident that some are aware and fixate strongly on the risk in the scene (e.g.~crossing vehicles or pedestrians), while others fixate fleetingly or, due to the presence of cognitive task modifiers, are unaware of the risk.}
    \label{fig_my_label}
    \vspace{-0.5cm}
\end{figure}

\subsection{Cognitive Task Modifiers}
\label{sec:data:conditions}
Although our dataset is collected in an in-lab experiment, we are still interested in subjects' behavior under a different task (in-car driving, engaged as well as distracted). We therefore included in our experiments several secondary cognitive task modifiers to artificially alter participant behavior in a way which might mimic certain variations in the behavior of drivers in real conditions. 
We aimed at modifiers that affected visual search patterns, but did not explicitly instruct the participants toward specific search targets. These modifiers affect the visual patterns by either changing the goal of visual search or by changing the scene appearance. The cognitive task modifiers were as follows:

\textbf{1. Null condition:} The subjects were given the task of supervising the driving of the car, looking for and flagging possible risky events or obstacles.

\textbf{2. Blurred condition:} Same as 1, but stimulus videos were blurred with a Gaussian kernel corresponding to $N\deg$ of visual field, making scene understanding and therefore the supervisory task harder, and affecting the visual search pattern.

\textbf{3. Vertical-flip condition:} Same as 1, but stimulus videos were flipped upside down, making scene understanding and therefore the supervisory task counter-intuitive, and affecting the visual search pattern.

\textbf{4. Road-only condition:} Same as 1, but subjects were asked to only fixate on road structure and not on dynamic obstacles such as cars and people.

\textbf{5. Reading-text condition:} Same as 1, but stimulus videos were overlaid with snippets of text of approximately even length at random image locations for $P$ seconds at an average interval of $Q$ seconds. Subjects were asked to read each text snippet while supervising the driving.

\subsection{Annotations}
\label{sec:data:annotations}
\begin{figure*}[t]
    \centering
    \includegraphics[width=0.9\linewidth,trim={0 9cm 0 16cm},clip]{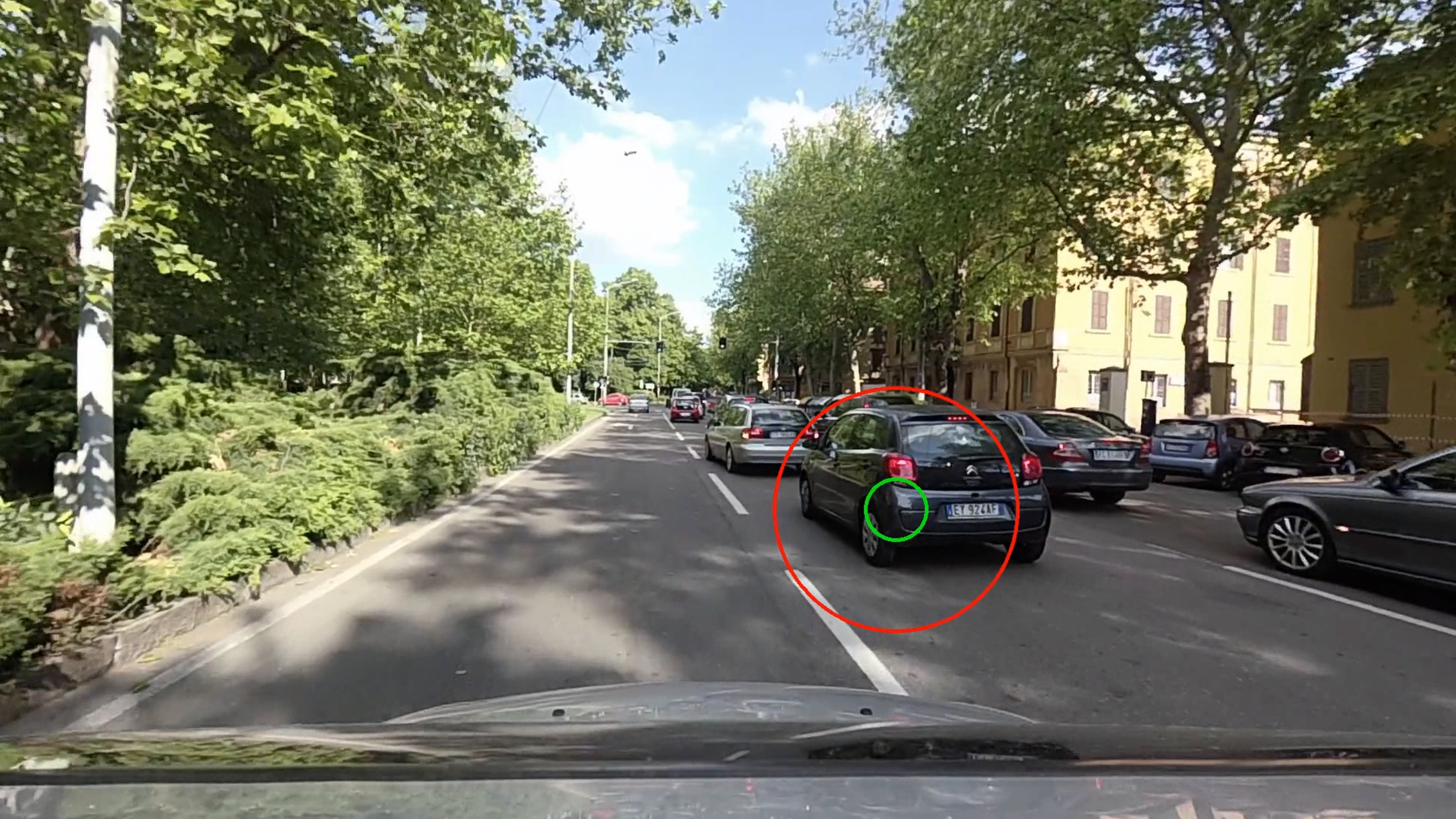}
    \includegraphics[width=0.9\linewidth,trim={0 9cm 0 16cm},clip]{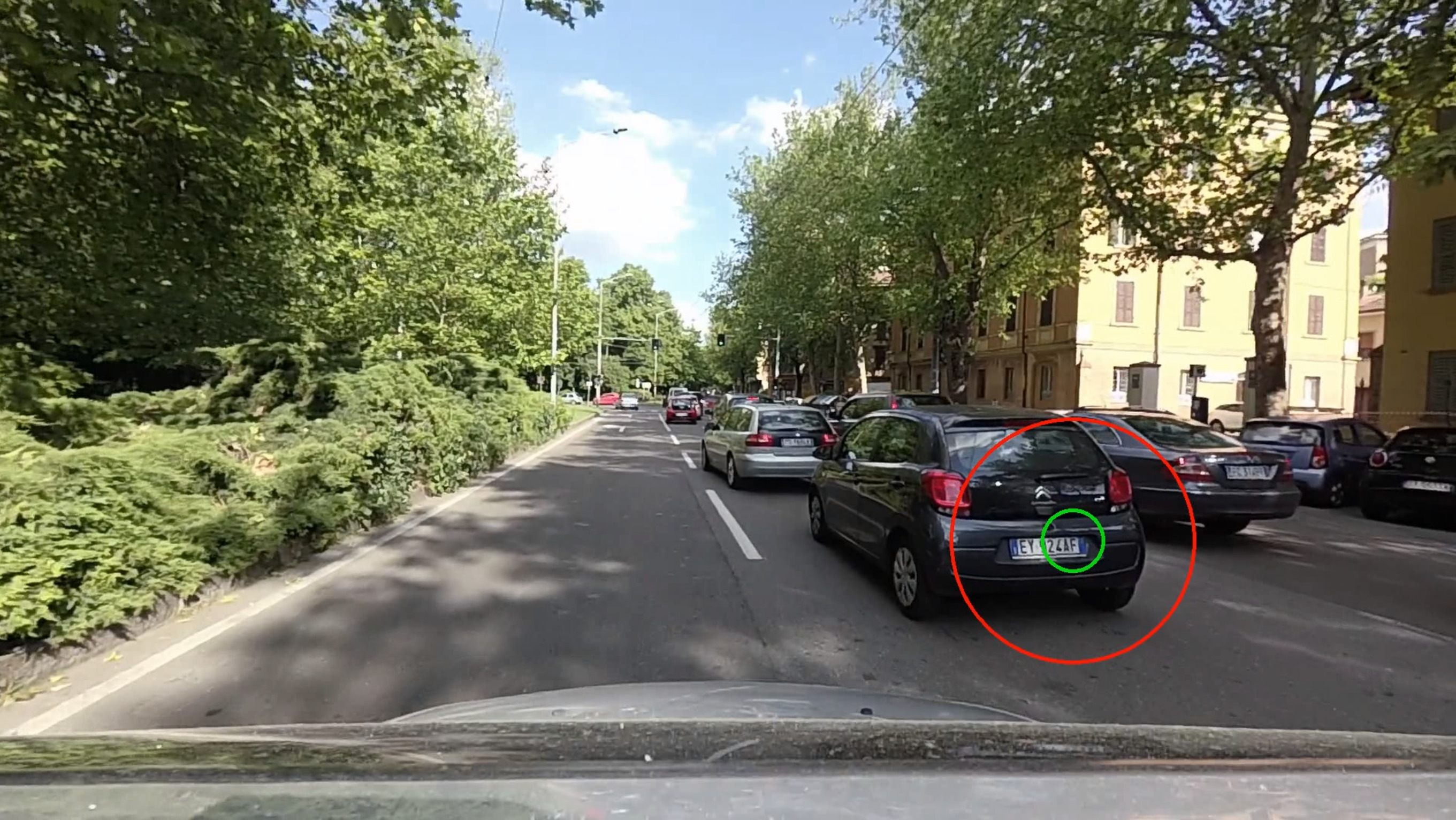}
    \includegraphics[width=0.9\linewidth,trim={0 9cm 0 16cm},clip]{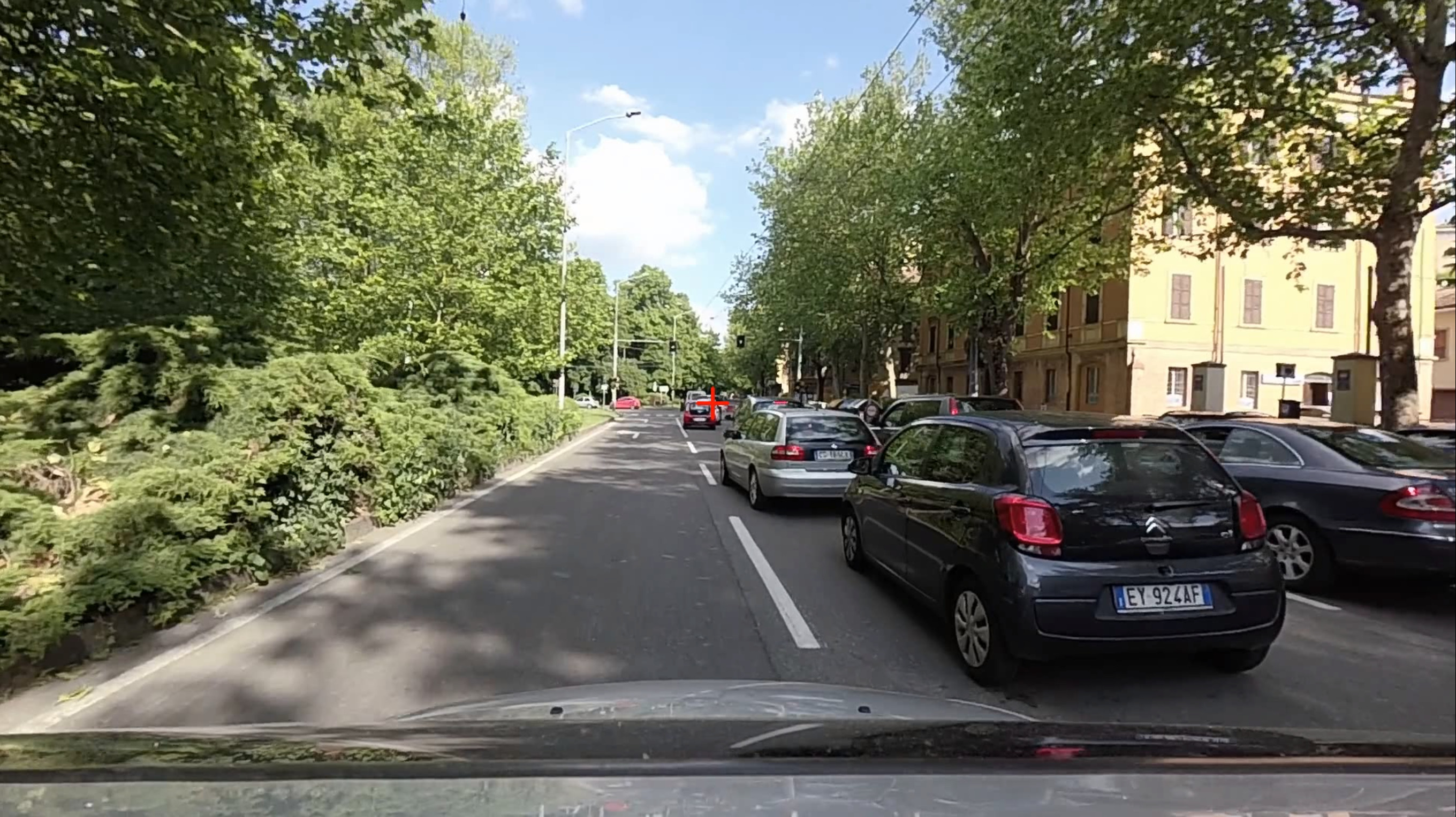}
    \caption{\textbf{Annotation example.} Left, center: Frames from the sequence with gaze estimates, along with the estimate of the central foveal vision area, and a larger red circle to mark what is probably outside the central vision area. Annotators were guided to treat these as approximations of the subject's attended gaze. Right: Annotators were asked to say whether the subject has attended to the red cross region and is aware of it. See the supp. mat. for additional details on the annotation.}
    \label{fig:annotation_example}
\end{figure*}

We annotated 53,983 sequences of approximately 10 seconds sampled randomly from within the data for attended awareness. 
While inferring attended awareness is difficult and subjective, and probing the subject's working memory directly is impossible, we devised a reproducible annotation scheme to explore a third person's estimate of a person's attended awareness. Our annotation protocol leverages the fact that humans are able to develop a theory of mind of other peoples' mental states from cues inherent in eye gaze~\cite{khalid2016eyes}. While the labels provided (as in any manually annotated dataset) are imperfect and have certain limitations, they are an important first step towards data-driven modeling of awareness. Annotators watched a video snippet where the subject's gaze was marked by two circles centered at the gaze point. One circle (green) size was set to the diameter of a person's central foveal vision area (2 degrees) at the viewing distance. Another circle (red) was set to a diameter four times the foveal vision circle. At the end of the video snippet, a random location was chosen and the annotators were asked whether they believe the subject has attended to that location on a scale between 1 and 5 (1-no, definitely not aware, 5-yes, definitely aware). Three different sampling types (object, edges, and non-objects) were used for sampling the final location. The annotations were linearly transformed to $[0, 1]$ in $L_A$ and provided a supervisory signal that the network (awareness predictor) tried to match. Annotators were asked whether the cursor corresponded to a well-defined object, whether they would expect to be aware of the location if they were driving and how surprised they by the appearance/behavior of the highlighted region. The annotations had good coverage across awareness levels and contained sufficient number of examples of both highly aware as well as unaware examples. Figure~\ref{fig:annotation_example} shows frames from an example annotation video. 
More annotation details and statistics in supp. material.

\section{Results}
\label{sec:experiments}
We now demonstrate the results from our model on several tasks of interest such as saliency, gaze refinement and attended awareness estimation. Our model, while applied here to a dataset which isn't strictly egocentric, could be straightforwardly extended to a true egocentric setting. 

\subsection{Saliency}
\label{sec:experiments:saliency}
In order to confirm that our model is capable of estimating visual saliency, we trained it on the DR(eye)VE dataset images and splits \cite{palazzi2018predicting}. As our approach assumes individual gaze samples per frame as the primary supervisory cue, we sampled gaze points from the fixation maps provided in the original dataset for every frame. We generated $p_G$, and compared it against the ground truth gaze map. For the standard measures of $KL$, cross-correlation and information gain (see \cite{palazzi2018predicting} for details), we obtained $1.734$, $0.565$, and $-0.0002$ respectively, comparing favorably to other algorithms tested on that dataset, such as \cite{cornia2016deep,wang2015consistent}. 

\subsection{Gaze refinement}
\label{sec:experiments:gaze_refinement}
In the following experiments, we show how our model is able to correct imperfect gaze estimates. In driving, driver gaze estimates can be obtained using camera-based driver monitoring systems (DMS). However, these estimates can be highly noisy due to the inherent process noise introduced during gaze tracking, or biased due to calibration errors in the DMS~\cite{kar2018performance}. Our hypothesis is that knowledge of where the person might look in the scene can help the model refine noisy and biased (miscalibrated) coarse estimates. We describe two experiments that were conducted to address these typical modes of failure.
For all experiments herein, we trained our model with a training dataset encompassing 8 urban day-time videos from our dataset with the highest subject and task coverage and adopted a fixed 80\%/20\% non-overlapping training/test split.

\subsubsection{Gaze Denoising}
For this experiment the gaze input to the gaze transform module is corrupted by noise, mimicking imperfect knowledge of a person's gaze. We use our model to try to identify the correct gaze point taking scene saliency into consideration. We use a spatially-varying additive white Gaussian noise model, where the input to the network $x_{\text{noisy}}$ is computed according to \cite{kellnhofer2019gaze360}:
\begin{align}
    x^{(i)}_{\text{noisy}}(t) =& x^{(i)}_{\text{true}}(t) + \eta^{(i)},\ \  \eta^{(i)} \sim \mathcal{N}(0,(\sigma^{(i)})^2),\\
    \sigma^{(i)} =& \max(\sigma^{(i)}_n, w*|x^{(i)}-x^{(i)}_0|) \nonumber
\end{align}
where $\eta$ is the additive noise and the standard deviation $\sigma^{(i)}$ increases as we get further from the center of the image along each coordinate $x^{(i)} \in \{x^{(1)},x^{(2)}\}$.
Our network denoises the gaze input by leveraging scene saliency information as encoded in the network. Fusing the noisy gaze location allows us to surpass the capability of a pure-saliency based model. The latter merely finds saliency image peaks that are close to the noisy gaze location and would be the straightforward way of incorporating saliency information. We also compare to an approach that relates gaze to the nearest object. In all cases we use a meanshift approach~\cite{comaniciu2002mean} to find nearby objects or peaks, with a standard deviation given by $\sigma_n\sqrt{\left(H^2 + W^2\right)}$, where $H$ and $W$ are the dimensions of the map. The results are summarized in Table~\ref{table_denoising} and demonstrate significant improvement with MAAD. 

\begin{table}[h!]
\centering
 \begin{tabular}{c| c c c c} 
 \hline
 Scenario & Raw & OBJ & SAL & MAAD \\ [0.5ex] 
 \hline
 $\sigma_n = 0.05$ & \textbf{15.7} & 62.0 & 25.7 &	18.1\\ 
 $\sigma_n = 0.10$ & 28.1 &	55.6 & 25.4	& \textbf{19.8}\\ 
 $\sigma_n = 0.15$ & 40.9 &	53.3 &	24.9 &	\textbf{21.0} \\ 
 $\sigma_n = 0.20$ & 51.7 &	53.1 &24.9 &\textbf{22.1} \\ \hline
 \end{tabular}
 \vspace{6pt}
 \caption{\label{table_denoising} Mean absolute error (in pixels) of noisy gaze recovery based on object attention: meanshift into nearby objects (OBJ), meanshift according to pure saliency map (SAL), and meanshift correction based on the MAAD gaze map, for different noise levels. Our approach improved upon other alternatives over a wide variety of input noise levels, far beyond the noise level at train time ($\sigma_n = 0.03$).}
 \vspace{-0.5cm}
\end{table}

\subsubsection{Gaze Recalibration}
\label{sec:experiments:recalibration}
We model imperfect calibration as an affine transformation. For this experiment, the DMS gaze input, $x_\text{noisy}$, to the gaze transform module is given by: 
\begin{equation}
    x_\text{noisy}(t) = T_{\text{correct}}\left(T_{\text{corrupt}}\left(x\right)\right)
\end{equation}
where $T_{\text{correct}},T_{\text{corrupt}}$ are both 2D affine transforms. We model the correcting transform, $T_{\text{correct}}$, as an MLP with one hidden layer.
The corruption transform is created by adding element-wise zero-mean Gaussian noise with standard deviation $\sigma^2_n$ to the transformation matrix and vector of an identity transform. 
We show the reduction in average error after calibration in Table~\ref{table_calibration}. By leveraging saliency information, we are able to naturally compensate for the calibration errors using the model.
\begin{table}[t]
\centering
 \begin{tabular}{c | c c} 
 \hline
 Scenario & Before & After \\ [0.2ex] 
 \hline
  $\sigma_n = 0.1$ & 0.23 & 0.13  \\ 
  $\sigma_n = 0.2$ & 0.58 & 0.16 \\ 
  $\sigma_n = 0.3$ & 1.12 & 0.33  \\ 
   $\sigma_n = 0.5$ & 1.23 & 0.59  \\ 
 \hline
 \end{tabular}
 \vspace{6pt}
 \caption{\label{table_calibration} The error was computed as the sum of the mean-squared error of the elements of the $T_\text{correct}$ and $T_\text{corrupt}^{-1}$ before and after recovery from miscalibration. For each noise level the errors were computed by averaging over eight optimization runs. 
 In Figure 4 (supp. material) we show the gaze map emitted by the model before and after recalibration with miscalibrated gaze input. }
\end{table}

\subsection{Driver Attended Awareness}
\label{sec:experiments:awareness}
In this experiment, we measure the model's ability to infer attended awareness. We do so by measuring the model's agreement with annotated third person estimate of attended awareness. We compare our approach to the following alternative of filtered gaze (FG) using a spatio-temporal Gaussian filter. We convolve each of the past gaze samples with a spatial Gaussian and utilize optic flow to temporally propagate the gaze information to the subsequent frames and aggregate them to form an awareness estimate. The optic flow mimics the subject's notion of object permanence under motion, and the spatial Gaussian account for subjective uncertainty accumulated over time as well as track limitations with optic flow. The results are given in Table~\ref{table_awareness}. 

\begin{table}[t!]
\centering
 \begin{tabular}{c | c c} 
 \hline
 Noise level & MSE, FG & MSE, MAAD \\ [0.5ex] 
 \hline
 $\sigma_n = 0.01$ & 0.357 & \textbf{0.138} \\ 
 $\sigma_n = 0.05$ & 0.359 & \textbf{0.135} \\
 $\sigma_n = 0.1$ & 0.425 & \textbf{0.138}  \\
 $\sigma_n = 0.15$ & 0.464 & \textbf{0.140}  \\
 \hline
 \end{tabular}
 \vspace{6pt}
 \caption{\label{table_awareness}Mean squared error awareness estimates with spatio-temporal Gaussian with optic flow (FG) and MAAD, as a function of input gaze noise level. Our approach significantly outperforms the baseline and is robust to the noise present in the gaze input.}
\end{table}
\begin{table}[b!]
\centering
 \begin{tabular}{c | c c} 
 \hline
 Ablation  & Awareness Estimate \\ [0.2ex] 
 \hline
   $\LLL_{ATT}$ & 0.262  \\ 
   $\LLL_{AA}$& 0.199  \\ 
   $\LLL_{T}$ & 0.159  \\ 
   $\LLL_{CON},\cdot$& 0.164  \\ 
   Full model &  \textbf{0.138}  \\ 
 \hline
 \end{tabular}
 \vspace{6pt}
 \caption{\label{table_ablations}Attended awareness estimation (mean squared error) on the test set using different ablations of MAAD. The testing noise level was set to be $\sigma_n =0.1$. More ablation results are presented in the supplementary material. }
\end{table}
\begin{figure}
    \centering
    \includegraphics[width=\linewidth]{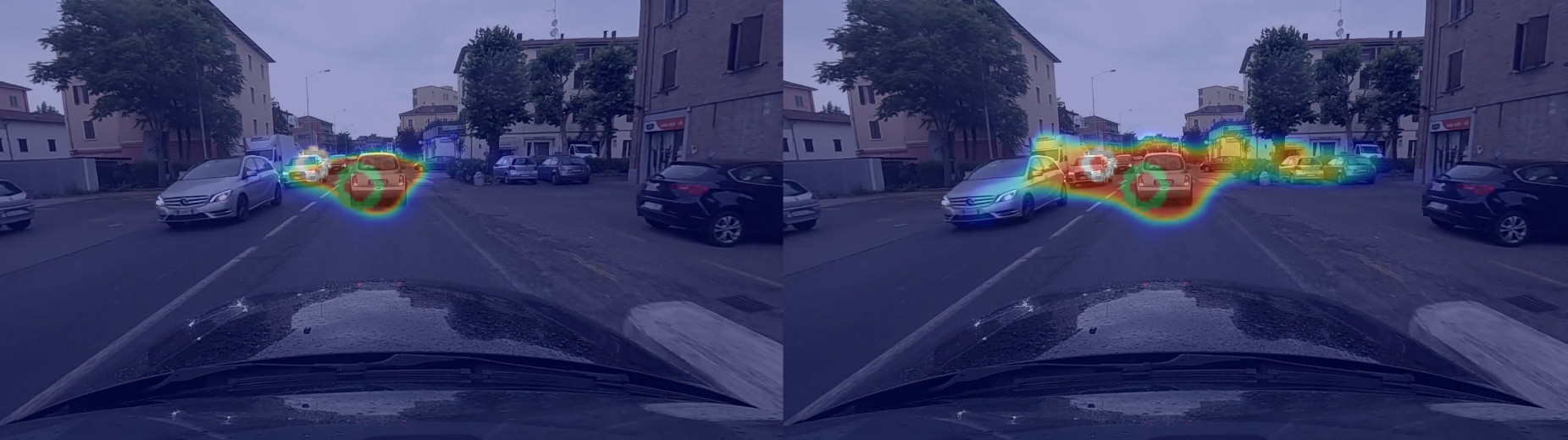}
    \includegraphics[width=\linewidth]{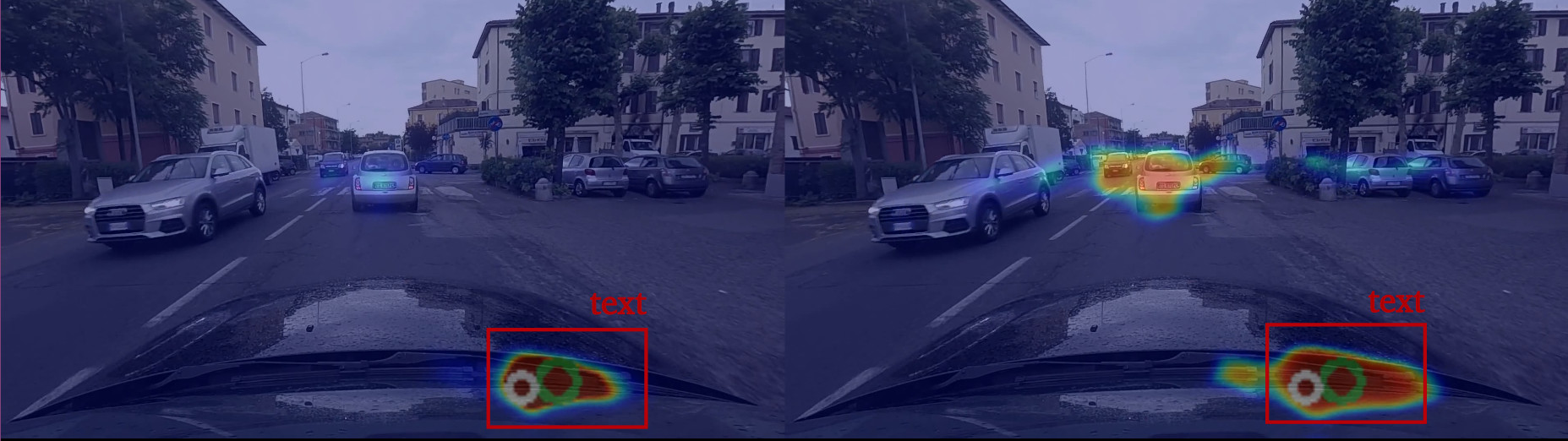}
    \includegraphics[width=\linewidth]{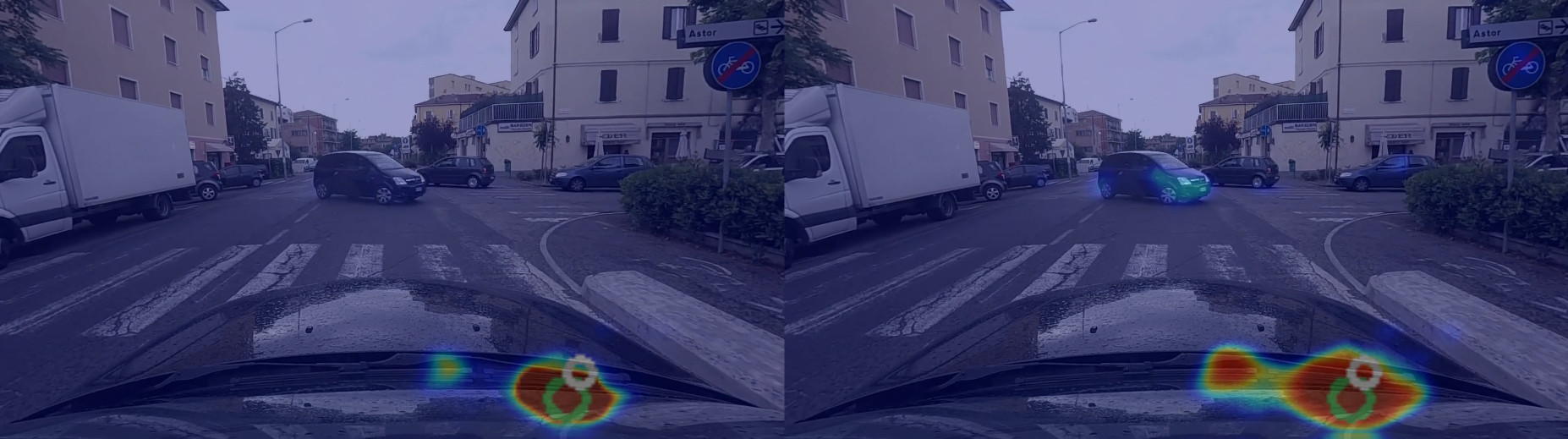}
    \caption{\textbf{Time evolution of gaze (left) and attended awareness (right) under the reading-text modifier condition.} Top: The subject gazes straight ahead and is aware of the car in front. Middle: The subject attends to text block and the gaze map shifts accordingly. The awareness map has two regions: one for the newly attended and the other for the previously attended location. Bottom: As the subject continues to read the text, the gaze map is localized in the bottom right, and the awareness map correctly captures how the subject is unaware of the scene in front.}
    \vspace{-0.5cm}
    \label{fig:awareness_figure2}
\end{figure}

\subsection{Ablation Experiments}
We performed a series of leave-one out ablations to investigate the impact of the various cost terms on the attended awareness estimation task. Both $\LLL_{ATT}$ and $\LLL_{AA}$ are crucial for more accurate awareness estimation (Table \ref{table_ablations}). 

\section{Discussion and Conclusion}
\label{sec:discussion}
We have introduced a new model which uses imperfect gaze information and visual saliency to reason about perceived attended awareness within a single model.
To train and validate our model, we generated a dataset which includes both high-accuracy gaze tracks as well as third person annotations for estimated attended awareness. MAAD can easily be extended to work with multi-image streams or scene graphs.
Although our subjects viewed pre-collected video stimuli as opposed to being part of a true egocentric vision-action loop, one advantage is that we could acquire multiple observations of the same video, enabling the study of \textit{distributions} rather than single scanpaths of attention for any given video. Our dataset can be compared in many ways to the related egocentric driving dataset from~\cite{palazzi2018predicting}.
Extending our work and model to study subjects who are in control of real vehicles is a topic for future work.

\section*{Supplementary Material}
In this supplementary material, we provide more details regarding the network modules, annotation dataset collection procedures and statistics, additional visualizations of network capabilities, and results from ablation experiments. 

\section{Model details}

\subsection{Network Architecture}

\subsubsection{Encoder Structure}
The encoder consists of 5 stacked layers of 2 different types of spatio-temporal convolutional modules. The first two layers (denoted as a `layer1' and `layer2') are taken from a pretrained ResNet18 structure. The weights of these two layers are frozen during training. The remaining 3 layers of the encoder (denoted as `S3D\_1', `S3D\_2', `S3D\_3') are separable 3D convolutional (S3D) modules. Each S3D module consists of two separate 3D convolutions, one for spatial and the other for temporal processing. Detailed structure of the separable convolutional encoder modules are shown in Table~\ref{tab:encoder_s3d_structure}. 
The output of `S3D\_3' undergoes a 3D convolution post-processing step in order to reduce the number of features from 512 to 128. The output of this post-processing step is then fed into the first decoder unit (DU5) of the decoder along with the side-channel information.

\subsubsection{Gaze Transform Module}
The gaze transform module consists of a single layer MLP whose output is encoded as a multi-channel Voronoi map which then is provided as a side channel input to the decoder units. The number of gaze points used per frame (for supervision as well as the side channel information) is fixed to be 3. The side-channel gaze input was corrupted by a spatially varying zero mean Gaussian white noise with  $\sigma = 0.0347$, to account for the uncertainty due to both the foveal center location and eye tracker error; both treated as two Gaussian independent sources. Each Voronoi channel encodes a particular distance related feature, such as $dx$, $dy$, $dx^2$, $dy^2$, $dxdy$, $\sqrt{dx^2 + dy^2}$. Additionally, we also provide a bit to encode whether a particular instance of the gaze input is dropped out (as a result of the dropout applied during training) and also whether the gaze value is a valid input or not (to indicate NaNs that occur in the gaze data primarily due to eye blinks and tracker error). The total number of channels for the gaze side information is 8.

\begin{table}[t]
    \centering
    \begin{tabular}{|c|c|}
    \hline
        Encoder S3D ID & Structure \\ \hline
        S3D\_1 & S3D(in=128, out=256)\\ \hline
        S3D\_2 & S3D(in=256, out=512) \\ \hline
        S3D\_3 & S3D(in=512, out=512)\\ \hline
    \end{tabular}
    \vspace{0.3cm}
    \caption{Detailed structure of the 3D convolution modules used in the Decoder Units. The spatial Conv3d in the encoder S3D modules uses kernel size of 1$\times$3$\times$3 and a stride length of 1. Similarly, the Conv3D responsible for temporal processing relies on a kernel of 3$\times$1$\times$1. A replication pad of size 1 is applied to the input before being processed by each of the Conv3D modules. }
    \label{tab:encoder_s3d_structure}
\end{table}
\begin{table*}[t]
    \centering
    \begin{tabular}{|c|c|c|}
    \hline
        Decoder Unit Id & Skip Module & Concatenated Module \\ \hline
        DU5 & NA & S3D(in=266, out=128) \\ \hline
        DU4 & NA & S3D(in=138, out=64)\\ \hline
        DU3 & NA & S3D(in=74, out=32) \\ \hline
        DU2 & S3D(in=128, out=128) & S3D(in=170, out=16)\\ \hline
        DU1 &  S3D(in=64, out=64) & S3D(in=90, out=16)\\ \hline
    \end{tabular}
    \vspace{0.3cm}
    \caption{Detailed structure of the S3D convolution modules used in the Decoder Units. The spatial Conv3d in the S3D modules uses kernel size of 1$\times$3$\times$3 and a stride length of 1. Similarly, the Conv3D responsible for temporal processing relies on a kernel of 3$\times$1$\times$1. A replication pad of size 1 is applied to the skip connection input to ensure that the output can be concatenated channel-wise to the other side-channel input and the previous decoder unit output.}
    \label{tab:du_structure}
\end{table*}\begin{table}[t]
    \centering
    \begin{tabular}{|c|c|}
    \hline
        Encoder S3D ID & Structure \\ \hline
        S3D\_1 & S3D(in=128, out=256)\\ \hline
        S3D\_2 & S3D(in=256, out=512) \\ \hline
        S3D\_3 & S3D(in=512, out=512)\\ \hline
    \end{tabular}
    \vspace{0.3cm}
    \caption{Detailed structure of the 3D convolution modules used in the Decoder Units. The spatial Conv3d in the encoder S3D modules uses kernel size of 1$\times$3$\times$3 and a stride length of 1. Similarly, the Conv3D responsible for temporal processing relies on a kernel of 3$\times$1$\times$1. A replication pad of size 1 is applied to the input before being processed by each of the Conv3D modules. }
    \label{tab:encoder_s3d_structure}
\end{table}
\begin{table*}[t]
    \centering
    \begin{tabular}{|c|c|c|}
    \hline
        Decoder Unit Id & Skip Module & Concatenated Module \\ \hline
        DU5 & NA & S3D(in=266, out=128) \\ \hline
        DU4 & NA & S3D(in=138, out=64)\\ \hline
        DU3 & NA & S3D(in=74, out=32) \\ \hline
        DU2 & S3D(in=128, out=128) & S3D(in=170, out=16)\\ \hline
        DU1 &  S3D(in=64, out=64) & S3D(in=90, out=16)\\ \hline
    \end{tabular}
    \vspace{0.3cm}
    \caption{Detailed structure of the S3D convolution modules used in the Decoder Units. The spatial Conv3d in the S3D modules uses kernel size of 1$\times$3$\times$3 and a stride length of 1. Similarly, the Conv3D responsible for temporal processing relies on a kernel of 3$\times$1$\times$1. A replication pad of size 1 is applied to the skip connection input to ensure that the output can be concatenated channel-wise to the other side-channel input and the previous decoder unit output.}
    \label{tab:du_structure}
\end{table*}

\subsubsection{Optic Flow Module}
The optic flow is provided as a 2-channel input, where the channels encode the flow in the horizontal and vertical direction respectively. We apply an adaptive average pool operator on the optic flow input to match the resolution of the decoder unit.

\subsubsection{Decoder Unit}
Each Decoder Unit (DU) can receive up to three sources of input, 1) the skip connections from the encoder, 2) the side channel information (gaze information, and optic flow) and 3) the output of the previous decoder unit, when available.

All S3D modules (for skip modules as well as side-channel+previous output modules) in each of the Decoder Unit uses a S3D unit with a kernel size of 1$\times$3$\times$3 for spatial and 3$\times$1$\times$1 for temporal processing. The input to each of the spatial and the temporal modules in the S3D uses a replication pad of size 1. An InstanceNorm3D and a ReLU nonlinearity is applied after the spatial and temporal processing. 

The output of the skip connection module is concatenated channel-wise to the side-channel information and the output of the previous decoder unit. The concatenated input is processed by another S3D module finally undergoes a bilinear upsampling to match the resolution size of the next decoder unit. The output of last decoder unit (DU1) undergoes a final bilinear upsampling stage to match the resolution of the size of model input (240$\times$135). The detailed structure of all the decoder units in the decoder is presented in Table~\ref{tab:du_structure}. 
 
The total number of channels from the side information is 10 (Voronoi gaze maps=$8$, and optic flow=$2$).
In general, the following relationship holds for the feature sizes:
\begin{equation*}
    n_{in}^{concat,DU(l)} = n_{out}^{skip,DU(l)} + n_{out}^{concat,DU(l+1)} + 10
\end{equation*}
where $n_{in}$ and $n_{out}$ are the number of input and output features respectively and $l \in [1,2]$ denotes the decoder unit id. For DU5, $n_{in} = n_{out}^{encoder,postproc} + 10$.

\subsubsection{Gaze and Awareness Convolutional Modules}
The output of the decoder is processed using a Conv2D with kernel operator of size 5$\times$5 and 6 output features to generate a feature map $M$. The 1D gaze heatmap ($p_G$) is produced from $M$ by a Conv1D operator with a kernel size of 1 followed by softmax operator to ensure that the heatmap is a valid probability distribution. Likewise, the awareness heatmap ($M_A$) is generated from $M$ by another Conv1D operator with a kernel size of 1 followed by a sigmoid operator to ensure that each pixel value remains between 0 and 1. Note that, the awareness map is NOT a probability distribution. 

\subsection{Cost Weights and Parameters}
Table \ref{tab:cost_coeffs} contains all the parameters and coefficients used for model training. These coefficients were chosen so that the relative magnitudes of the different supervisory terms were comparable. The regularization terms are roughly an order of magnitude lower than the supervisory cues. The gaze and awareness supervision costs are computed only on valid gaze points (gaze points that are not NaNs).

\begin{table}[h]
    \centering
    \begin{tabular}{|c|c|}
        \hline
       $\alpha_{\text{G}}$ & $1.2$ \\ \hline
       $\alpha_{\text{ATT}}$  & $12.0$ \\ \hline
       $\alpha_{\text{AA}}$ & $1.0$ \\ \hline
        $\alpha_{\text{S-A}} $ & $100.0$ \\ \hline
        $\alpha_{\text{S-G}} $ & $5\times 10^{10}$ \\ \hline
        $\alpha_{\text{T}}$  & $600.0$ \\ \hline
        $\alpha_{\text{DEC}} $ & $1.5\times 10^{6}$ \\ \hline
        $\alpha_{\text{CAP}} $ & $0.01$ \\ \hline
        $\alpha_{\text{CON-G}} $ & $1\times 10^{7}$ \\ \hline
        $\alpha_{\text{CON-A}} $ & $10.0$ \\ \hline
        $w_{OF}$ & $0.5$ \\ \hline
        $\epsilon_{\text{DEC}} $ & $0.2$ \\ \hline
    \end{tabular}
    \vspace{0.3cm}
    \caption{Cost term coefficients and parameters used for training. }
    \label{tab:cost_coeffs}
\end{table}
\begin{table}[t!]
    \centering
     \begin{tabular}{c c c} 
     \toprule
    Modifier & Num. annotations & Mean awareness \\ [0.5ex] 
     \midrule
     Null & 16,366 & 0.719 \\ 
     Blurred & 8,346 & 0.657  \\ 
     Flipped & 8,311 & 0.674 \\ 
     Road-only & 9,665 & 0.542 \\ 
     Reading-text & 11,295 & 0.593 \\
     \bottomrule
     \end{tabular}
     \vspace{6pt}
     \caption{\label{table_annotations} Number of annotations and mean awareness from annotations grouped according to cognitive task modifier. Annotations reflect the variability in  awareness of locations under certain cognitive modifiers, including conditions where we expect reduced awareness of annotated locations (e.g. reading text and road-only conditions).}
\end{table}

\begin{figure*}
    \centering
    \includegraphics[width=0.49\linewidth]{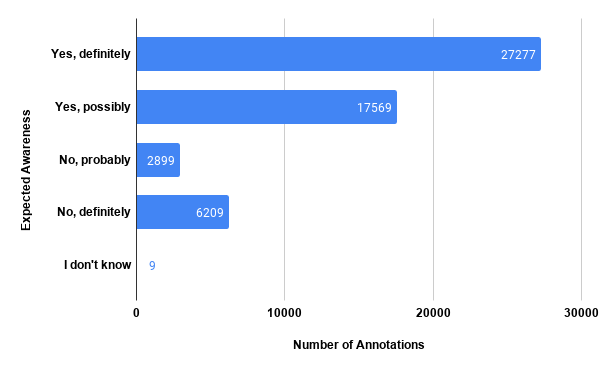}
    \includegraphics[width=0.49\linewidth]{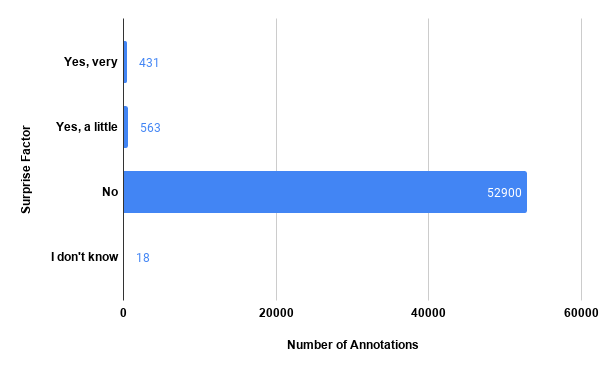}
    \includegraphics[width=0.49\linewidth]{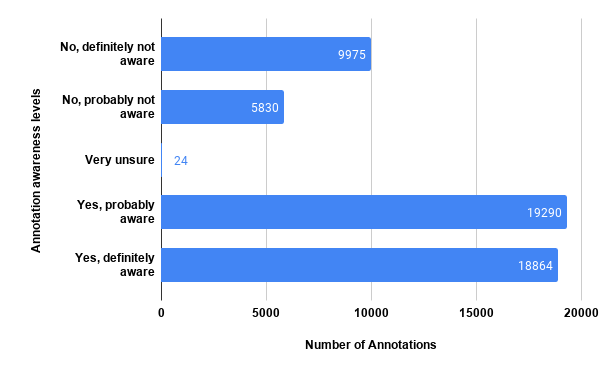}
    \includegraphics[width=0.49\linewidth]{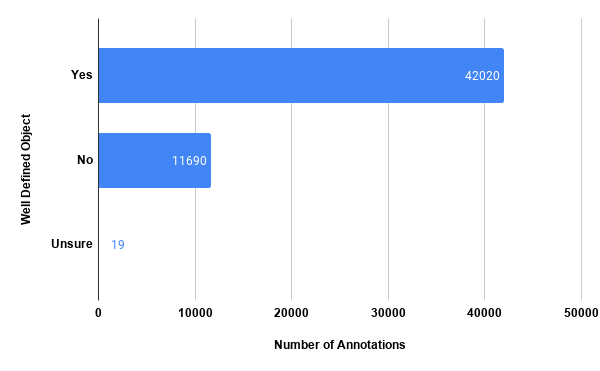}
    \caption{Annotator response distribution. Very few awareness labels were truly unsure (probably due to the explicit definition of the questions).}
    \label{fig:response_distribution}
\end{figure*}
\section{Annotation Dataset Details}
\begin{table}[h]
    \centering
    \begin{tabular}{|c|c|c|}
        \hline
        VIDEO ID & Time & Weather \\ \hline
        VID06 & Morning & Sunny \\ \hline
        VID07 & Evening & Rainy \\ \hline
        VID10 & Evening & Rainy\\ \hline
        VID11 & Evening & Cloudy \\ \hline
        VID26 & Morning & Rainy\\ \hline
        VID35 & Morning & Cloudy\\ \hline
        VID53 & Evening & Cloudy \\ \hline
        VID60 & Morning & Cloudy \\ \hline
    \end{tabular}
    \vspace{0.3cm}
    \caption{Time of the day and the weather condition for each of the videos (from the Dr(Eye)ve dataset) used for MAAD training. All driving sequences occur in an urban (downtown) setting. }
    \label{tab:metadata_vids}
\end{table}

Table~\ref{table_annotations} shows the breakdown of the labelled set. Table~\ref{tab:metadata_vids} contains information regarding the time of the day and the weather condition for all the 8 video sequences (from the Dr(Eye)ve dataset) we used for MAAD model training. 

We randomly sampled approximately 10s clips from these 8 videos from within the data we collected for third-party attended awareness annotation. The gaze data was overlaid on the video clip and in the last frame of the clip a random location was chosen and marked with a red cross. This random location was chosen equi-probably from objects, edges or anywhere in the image. After the annotators watched the video, they were asked whether they believed the subject had attended to the location marked with the red cross. More specifically, the annotators answered the following questions:
\begin{itemize}
    \item Do you think the driver is aware of the object/area? (red cross; must be near the green circle at some point in the video, not being near at the end of the video is fine, if it is close and moving along with the object, we want a human judgment of someone who has the extra knowledge and is focusing on this)
    a) Yes, definitely aware b) Yes, probably aware c) Very unsure d) No, probably not aware e) No, definitely not aware
    \item Is the red cursor on a well-defined object such as a car or person? (not well defined: exit, piece of road, something you cannot put a boundary around. If it is part of an object, then it is still well defined. For example, building is not well defined because it’s a large area and cannot be separated from the ground) 
    a) Yes b) No c) Unsure.
    \item If you are driving and are concentrated on driving, would you expect to be aware of this object? (red cross. Based on everything you see in the video)
    a) Yes, definitely b) Yes, possibly c) No, probably d) No, definitely e) I don't know.
    \item Were you surprised by the behavior or appearance of the highlighted object/region in the video? (red cross; jumped suddenly, didn’t expect to see it, didn’t see it coming, near accidents)
    a) Yes, very b) Yes, a little c) No d) I don't know.
    
\end{itemize}

\begin{figure}
    \centering
    \includegraphics[width=0.49\linewidth]{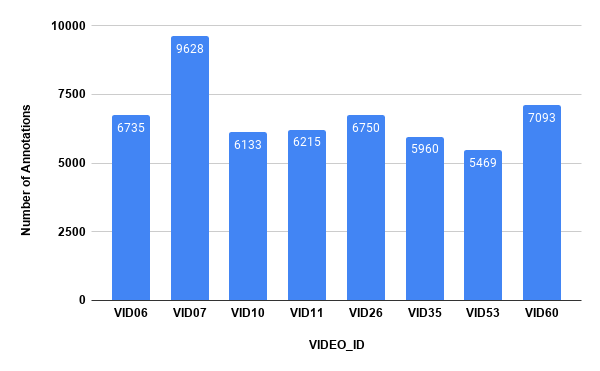}
    \includegraphics[width=0.49\linewidth]{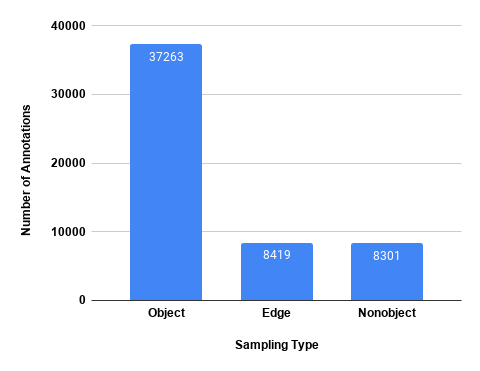}
    \caption{Top: Distribution of annotation video snippets. All video sequences were well represented, aside from some variance due to the criteria used to select desired and undesired sampling points. The video ids on the horizontal axis refer to the original video sequence ids in the Dr(Eye)ve dataset. Bottom: Distribution of final location for annotation according to sampling type. }
    \label{fig:distribution}
    
\end{figure}

Figures~\ref{fig:response_distribution} and \ref{fig:distribution}  the responses from the annotators and the distribution of annotation video snippets respectively. 

\section{Examples of Calibration Optimization}
Figure~\ref{fig:calibration} shows more examples of how the network successfully corrects a miscalibrated side-channel gaze input. In each of the examples in the figure, before correction the miscalibrated gaze distorts the heatmap and pulls it away from the ground truth gaze. As the networks learns the correction transform (for this experiment, the correction transform was learned by training the network on the test split that was used during the original model training phase), it corrects for the miscalibration and the heatmap begins to align closely with the ground truth gaze. Note that, during the optimization procedure for learning the correction the weights of the entire network except that of $T_{correct}$ are kept frozen. 

\section{Visualization of Denoising Mean Shift Traces}
The meanshift algorithm is a procedure for locating the local maxima---the modes---of a density function. For the gaze denoising experiment, we perform the exact same meanshift procedure on three different density maps a) the gaze-conditioned saliency map, b) pure saliency map and c) the mask image that encodes the objects in the scene. 
Figure~\ref{fig:meanshift} shows different examples of the traces of the mean shift procedure on the mask image, gaze map without and with side channel gaze. In general, we see that when mean shift is performed on the gaze-conditioned saliency maps the resulting mode is closer to the ground truth (right-most column in Figure~\ref{fig:meanshift}).

\section{Ablation Experiments}
\begin{table}[t]
\centering
 \begin{tabular}{c | c} 
 \hline
 Ablation  & Awareness Estimate \\ [0.2ex] 
 \hline
   $\LLL_{ACAP}$ & $0.167$  \\ 
   $\LLL_{DEC}$& \textcolor{red}{$0.073^{*}$}  \\ 
   $\LLL_{S-A}$ & $0.157$ \\ 
   $\LLL_{S-G}$& $0.143$  \\ 
   $\LLL_{ACAP}, \LLL_{AA}$& \textcolor{blue}{$0.138^{*}$} \\ 
   $\LLL_{ACAP},\LLL_{S-G}$& $0.270$ \\ 
   $\LLL_{ACAP},\LLL_{ATT}$& $0.444$ \\
   $\LLL_{ACAP},\LLL_{T}$& $0.165$ \\
   $\LLL_{ACAP}, \LLL_{S-G},\LLL_{S-A}$& \textcolor{blue}{$0.134^{*}$} \\
   $\LLL_{T},\LLL_{S-G}$& $0.264$ \\
   $\LLL_{T},\LLL_{S-G},\LLL_{S-A}$& $0.146$ \\ 
   Full model &  \textbf{0.138}  \\ 
 \hline
 \end{tabular}
 \vspace{6pt}
 \caption{\label{table_ablations_supp}Attended awareness estimation (mean squared error) on the test set using different ablations of MAAD. The testing noise level was set to be $\sigma_n =0.1$. The result highlighted in red indicates the anomalous case in which the awareness heatmap is no longer spatially localized and hence results in gross overestimation of attended awareness. The results highlighted in blue indicate ablations for which the results were comparable to the full model but resulted in training instability. For more discussion on results with asterisk please refer to the text in Section 5. }
\end{table}

We performed a set of leave-$N$-out ablations to investigate the impact of different regularization terms on the network's ability to estimate attended awareness. Table~\ref{table_ablations_supp} shows the mean squared error in the awareness estimation for different ablations that we tested. 

\textbf{Regularization for stability}: One of the key functions of the regularization terms is to provide stability during training. In our ablation experiments we found that ablating the attention capacity regularization term ($\LLL_{ACAP}$) in general, resulted in training instability and in truncated training runs despite seemingly comparable (and at times better) awareness estimation scores to the full model. 

We also experimented with a different network architecture in which the S3D modules in the decoder units were replaced with standard Conv3D modules. Due to the larger number of parameters for Conv3D modules, the number of layers in the encoder and decoder were reduced to 4. 
For these architectures,  we found that including the spatial regularization for the gaze map ($\LLL_{S-G}$) was critical for stability during training. 

In general, from our ablation experiments we recommend that for both the S3D and non-S3D versions of the model, the spatial regularization ($\LLL_{S-G}$) and the attention capacity ($\LLL_{ACAP}$) cost terms should be added to improve training stability.

\begin{figure}[b]
    \centering
    \includegraphics[width=0.49\linewidth]{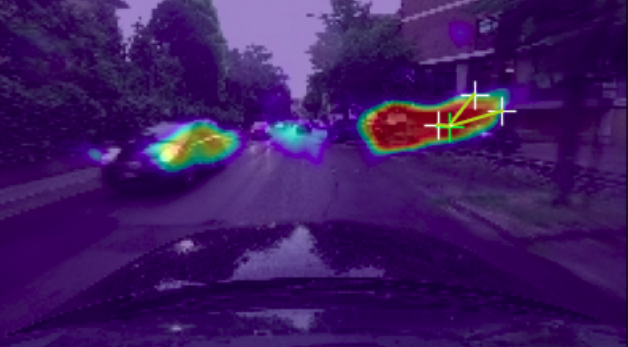}
    \includegraphics[width=0.49\linewidth]{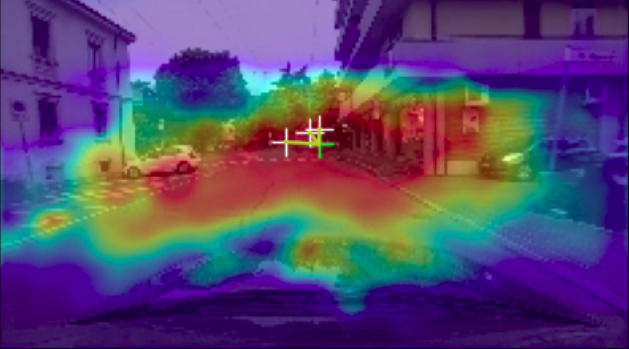}
    \caption{Comparison of awareness heatmap with (left) and without $\LLL_{DEC}$ (right). The heatmap is much more localized when decay term is present in the cost function.}
    \label{fig:dec_ablation}
    
\end{figure}

\begin{figure*}
    \centering
    \includegraphics[width=\linewidth]{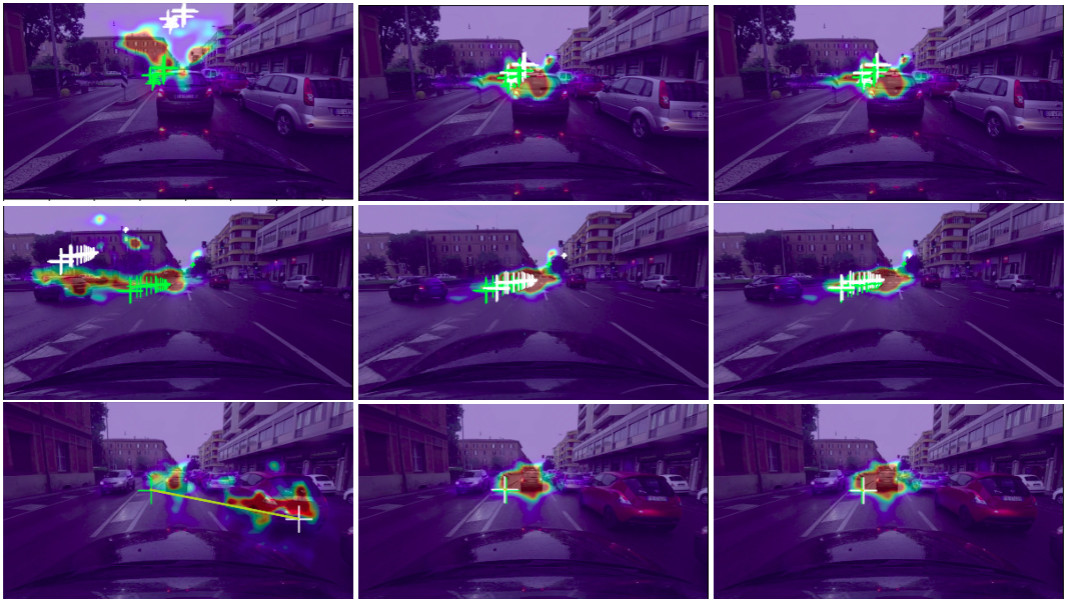}
    \caption{Examples of how the network learns a correction transform to correct for a miscalibrated gaze input. In each of the examples, the corrupted gaze input is marked as white cross hairs and the ground truth gaze is marked as lime-green cross hairs. Left column: Gaze maps before calibration. Due to the corruption applied, we can see that the side channel gaze is far away from the ground truth gaze. Middle and Right Column: As the optimization progresses, the networks learns to correct for the miscalibration and brings the side channel input close to the ground truth. In all these examples, the noise level was set to be 0.3.}
    \label{fig:calibration}
\end{figure*}

\begin{figure*}
    \centering
    \includegraphics[width=\linewidth]{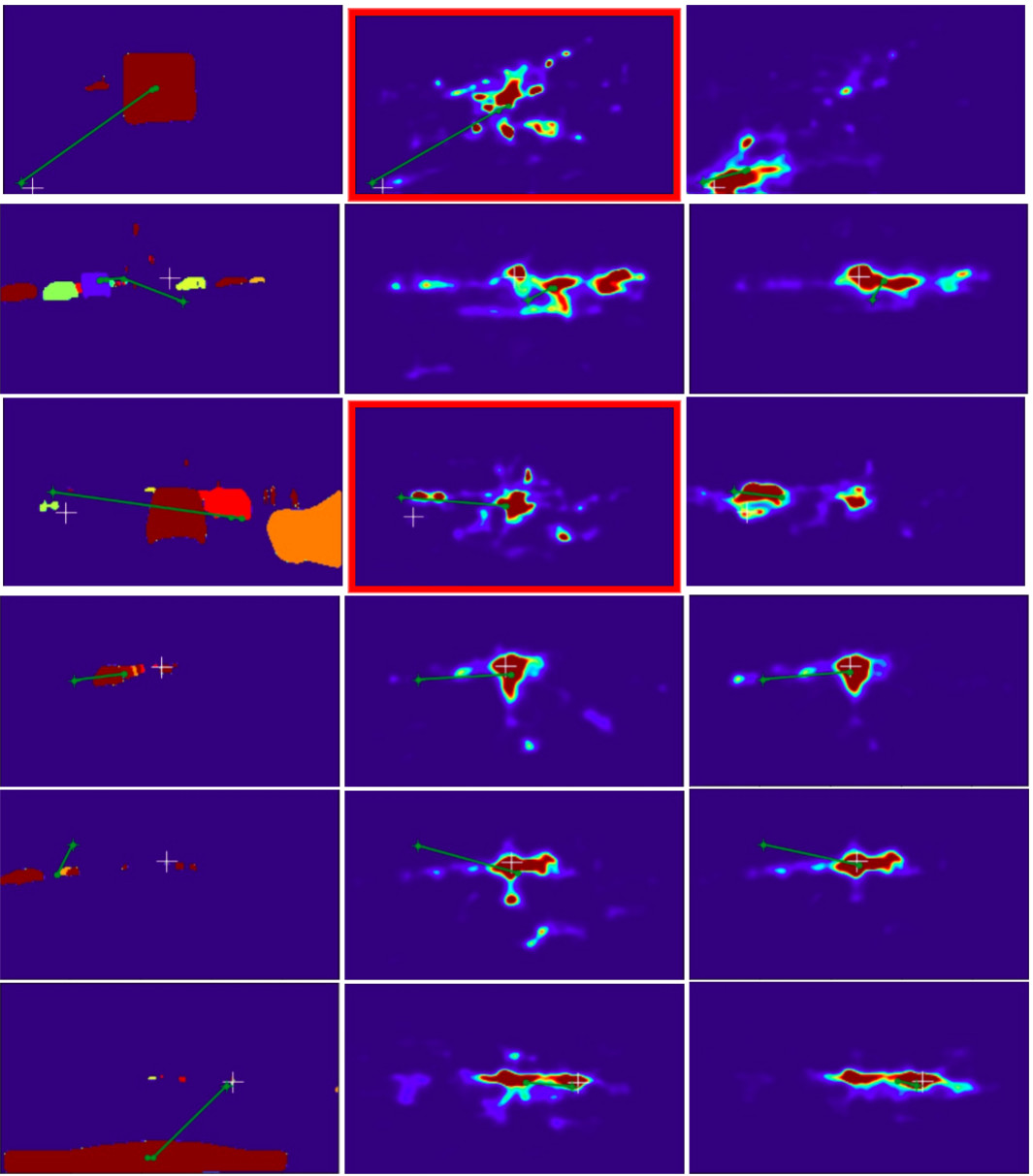}
    \caption{Examples of meanshift operation performed on Left: Object Masks, Middle: Gaze map without side-channel (pure saliency) information. The examples highlighted using the red rectangle indicate extreme failure cases. Right: Gaze map with side channel noisy gaze (our approach). The meanshift sequences is shown as green polylines. The starting point (the noisy gaze) of the sequence is indicated using a green crosshair. The ground truth gaze is denoted as white cross on the images. Our approach with noisy side channel gaze outperforms the object-based and pure saliency-based approaches. }
    \label{fig:meanshift}
\end{figure*}

\textbf{$\LLL_{DEC}$ ablation}: Although removing the decay term, ($\LLL_{DEC}$), resulted in better awareness estimation scores (row 2, Table~\ref{table_ablations_supp}, this was due to the fact that without $\LLL_{DEC}$ the awareness heatmap was no longer spatially localized as shown in Figure~\ref{fig:dec_ablation} essentially resulting in over-estimation of attended awareness. Over-estimation of driver awareness (model falsely predicting that the driver is aware of something when they are not) can lead to undesirable consequences when used in safety warning systems in autonomous vehicles. Additionally, utilizing $\LLL_{DEC}$ also accelerated the convergence of the model during training.

\section{Influence of Cognitive Task Modifiers}
During the dataset collection procedure we opted for a high-accuracy gaze tracker. However, this raises a question about the effect of the cognitive task modifier in a passive observation experiment.
In order to investigate the impact of cognitive task modifiers as a latent factor that could influence awareness estimation accuracy, we trained MAAD exclusively on training data collected under the `null condition'. This model was then evaluated on the data collected under the remaining cognitive task modifier conditions. 

From Table~\ref{table_task_modifier} we can see that a model that was trained exclusively on null condition data performed worse on the other task modifiers compared to the full model. However, as shown in Table~\ref{table_awareness_nullcondition}, the null condition model still did considerably better than the spatio-temporal Gaussian baseline (FG) with optic flow. These results indicate that the model is sensitive to the cognitive task that the subject is executing.
Future work will explore ways to disentangle this latent factor within the network capabilities. 

\begin{table}
\centering
 \begin{tabular}{c | c | c} 
 \hline
 Task Modifier & Null Condition Model & Full Model \\ [0.2ex] 
 \hline
   Null Condition & \textbf{0.110} & 0.139 \\
   Reading-Text & 0.211  & \textbf{0.132}  \\ 
   Blurred & 0.220 & \textbf{0.140}  \\ 
   Flipped & 0.231 & \textbf{0.166}  \\ 
   Roadonly & 0.171 & \textbf{0.114} \\ 
 \hline
 \end{tabular}
 \vspace{6pt}
 \caption{\label{table_task_modifier}Awareness estimation mean squared error: Breakdown of the results according to cognitive task modifier type for full model and the model training on only null condition data. }
\end{table}

\begin{table}[t]
\centering
 \begin{tabular}{c | c c} 
 \hline
 Noise level & MSE, MAAD &  MSE, FG\\ [0.5ex] 
 \hline
   Null Condition & \textbf{0.110} & 0.492 \\
   Reading-Text & \textbf{0.211}  & 0.393 \\ 
   Blurred & \textbf{0.220} & 0.425  \\ 
   Flipped & \textbf{0.231} & 0.454  \\ 
   Roadonly & \textbf{0.171} & 0.330 \\ 
 \hline
 \end{tabular}
 \vspace{6pt}
 \caption{\label{table_awareness_nullcondition}Mean squared error awareness estimates with spatio-temporal Gaussian with optic flow (FG) and our proposed approach (MAAD) according to cognitive task modifier type for the model trained only on null condition data.}
\end{table}
\section{Examples of Gaze and Awareness Maps}
\begin{figure*}
    \centering
    \includegraphics[width=\linewidth]{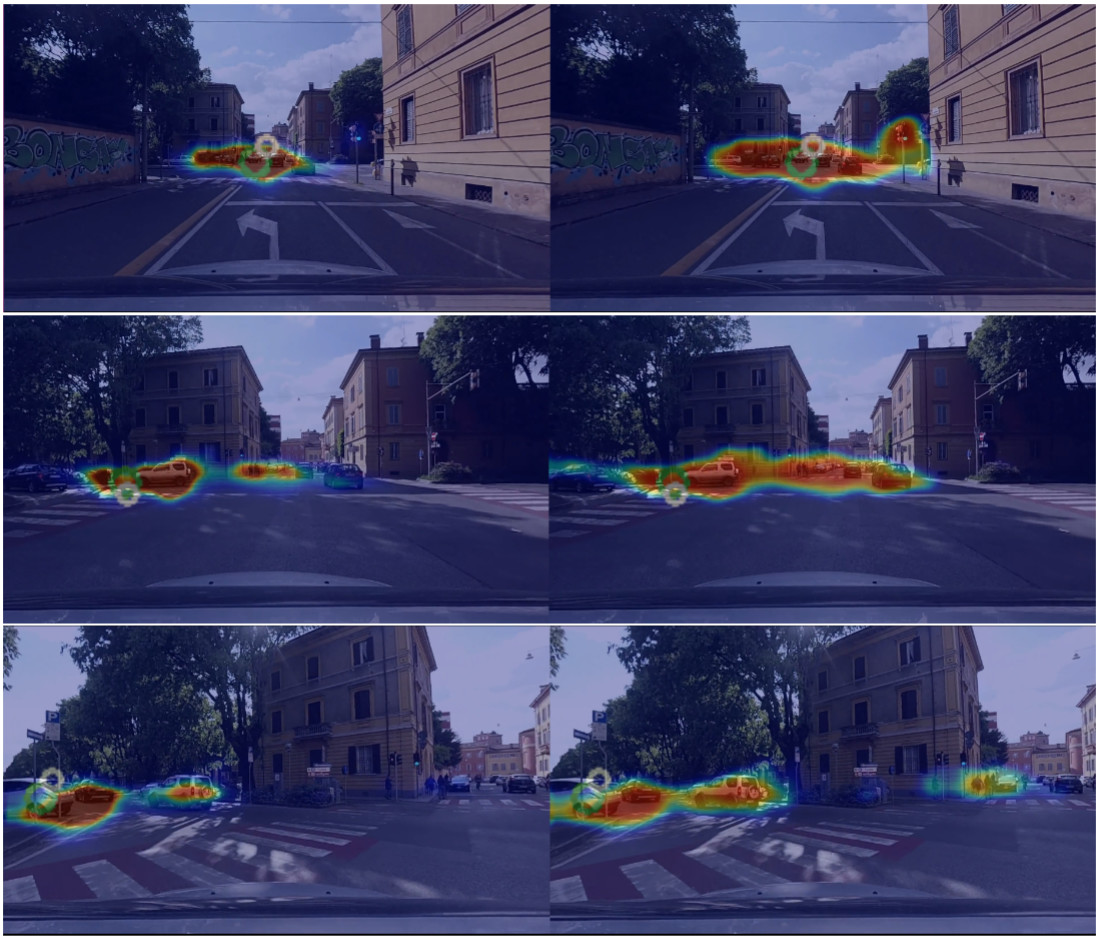}
    \caption{Gaze and awareness map as the ego car performs a left turn. Top Left: The car approaches an intersection and is about to make a left turn. The gaze map and the awareness map are primarily concentrated in the center of image. Middle: The car has started the left turn maneuver. The gaze map has started to shift leftward. However, the awareness map is much more smooth and indicates awareness of the cars ahead of the ego car in the previous frame. Bottom Middle: The left turn maneuver is almost complete and the gaze map is almost completely shifted to the left hand side. The awareness map still exhibits temporal persistence of objects that were attended to a few seconds before. In this figure, the gaze and awareness maps are in the left and right column respectively. The green circle indicate the ground truth gaze and the white circles indicate the noisy side channel information fed into the network during inference. }
    \label{fig:gaze-awareness-1}
\end{figure*}

\begin{figure*}
    \centering
    \includegraphics[width=\linewidth]{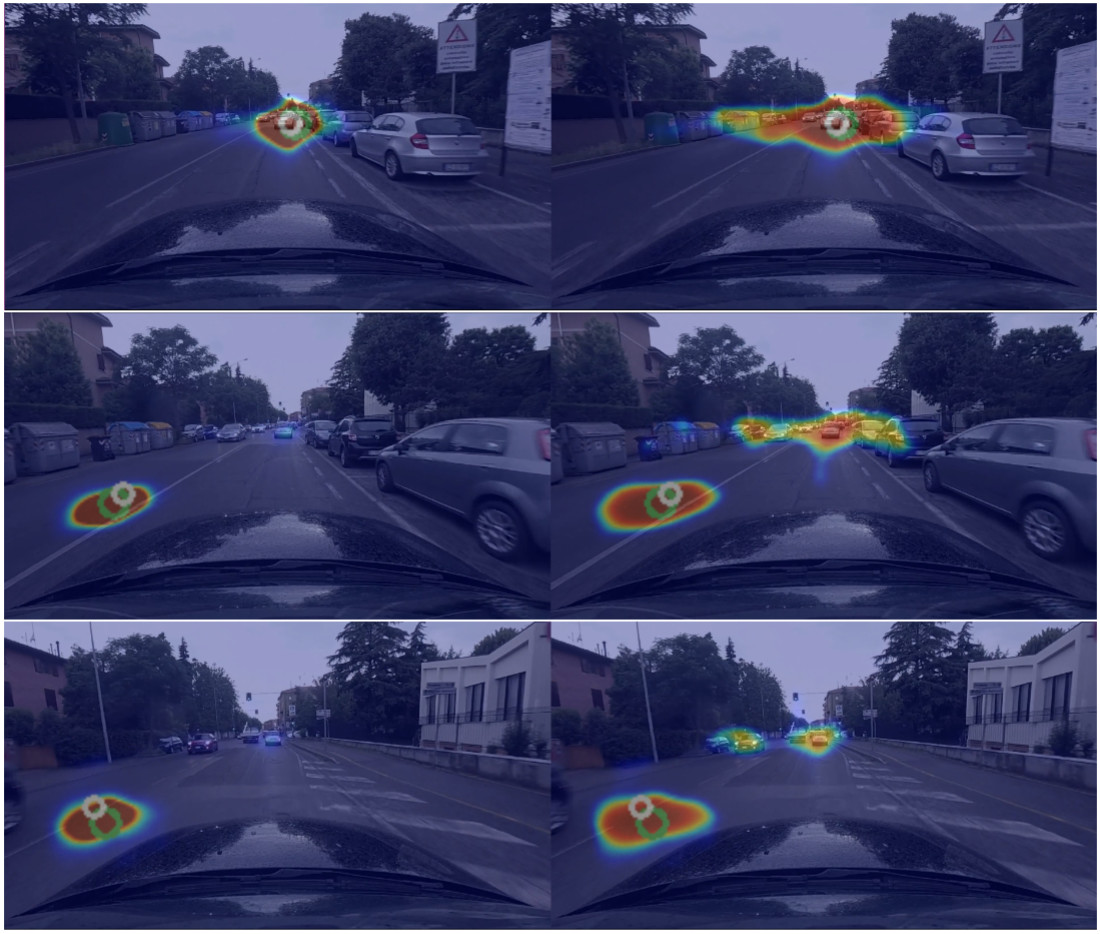}
    \caption{Gaze conditioned saliency and awareness maps with shifting gaze and incoming traffic: Top: The gaze is fixated straight ahead and on the incoming traffic. Middle: The gaze map shifts to the bottom left (reading text). The awareness map exhibits multiple regions of activation: a) for the newly attended region, b) the car straight ahead and c) the incoming traffic. Bottom: The gaze maps remains almost the same as the subject continues to gaze in the bottom left. As the incoming traffic approaches closer to the ego car, the activation levels of the awareness map have weakened and furthermore, the activation regions have separated indicating spatial and temporal persistence attached to objects.}
    \label{fig:gaze-awareness-2}
\end{figure*}

\begin{figure*}
    \centering
    \includegraphics[width=\linewidth]{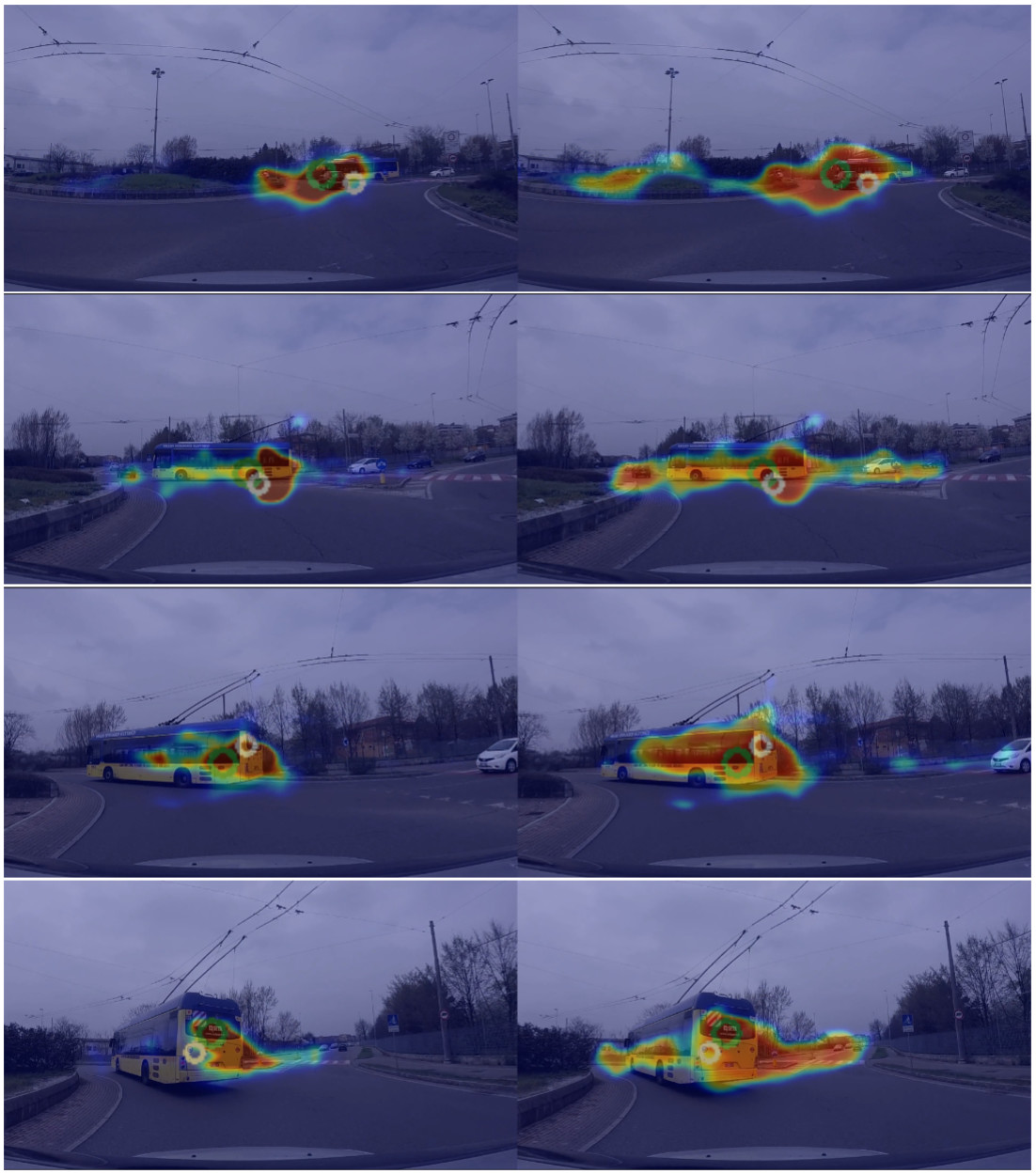}
    \caption{Gaze conditioned saliency and awareness maps with occlusion: Top: As the ego car approaches the traffic roundabout, the gaze is impinged on the car ahead. The awareness map reflects the fact that the subject is aware of the car as well. Top Middle: The gaze has shifted toward the bus on the right and the car ahead is about to be occluded by the bus. The awareness is split between the car and the vehicles on the right. Bottom Middle: The car is no longer visible due to occlusion. The gaze and the awareness activation is solely on the bus. Bottom: The car has reappeared in the visual field after occlusion. The gaze activation continues to be on the bus. The awareness map reassigns positive awareness on the previously attended car demonstrating how the model can `remember' past attended objects despite occlusions.}
    \label{fig:gaze-awareness-3}
\end{figure*}

Figures~\ref{fig:gaze-awareness-1},\ref{fig:gaze-awareness-2} and \ref{fig:gaze-awareness-3} are more examples of gaze and awareness maps for different interesting scenarios that arise during driving. Figures~\ref{fig:gaze-awareness-1} and \ref{fig:gaze-awareness-2} are examples of gaze conditioned saliency and highlights the spatio-temporal persistence of awareness in situations where the subject's gaze shifts between multiple driving-relevant entities (such as pedestrians, traffic lights and other cars) in the scene. Figure~\ref{fig:gaze-awareness-3} is a unique example which illustrates how the network can handle occlusions. 

\end{document}